\documentclass[preprint, 10pt]{aastex}    
\singlespace

\usepackage{flushrt}
\usepackage{graphicx}
\DeclareGraphicsExtensions{.eps,.jpg,.pdf,.mps,.png}

\begin{document}
\slugcomment{The Astrophysical Journal, 548:\ 564-584, 2001 February 20}
\shortauthors{Willick \& Batra}
\shorttitle{$H_0$ From Cepheids and Flow Model}

\newcommand{\be}{\begin{equation}}
\newcommand{\ee}{\end{equation}}
\newcommand{\velmod}{{\small VELMOD\/}}
\newcommand{\omegam}{\Omega_{\rm m}}
\newcommand{\omegal}{\Omega_\Lambda}
\newcommand{\vev}[1]{\left\langle#1\right\rangle}
\newcommand{\bfv}{\mathbf v}
\newcommand{\bfr}{\mathbf r}
\newcommand{\bfk}{\mathbf k}
\newcommand{\bfp}{\mathbf p}
\newcommand{\bfw}{\mathbf w}
\newcommand{\gtsima}{$ \buildrel > \over \sim \,$}
\newcommand{\ltsima}{$ \buildrel < \over \sim \,$}
\newcommand{\simgt}{\lower.5ex\hbox{\gtsima}}
\newcommand{\simlt}{\lower.5ex\hbox{\ltsima}}
\newcommand{\infint}{\int_{-\infty}^\infty \! }
\newcommand{\zinfint}{\int_{0}^\infty \! }
\newcommand{\hmpc}{h^{-1}\,{\rm Mpc}}
\newcommand{\kms}{\ifmmode\,{\rm km}\,{\rm s}^{-1}\else km$\,$s$^{-1}$\fi}
\newcommand{\degs}{\ifmmode^\circ\else$^\circ$\fi}
\newcommand{\kmsmpc}{\kms\,{{\rm Mpc}}^{-1}}
\hyphenation{an-is-o-tro-pies}
\hyphenation{sam-ples}
\hyphenation{an-is-o-tro-py}
\newcommand{\etal}{et al.} 
\def \arcmin{\hbox{$^\prime$}} 
\def \farcs{\hbox{$.\!\!^{\prime\prime}$}} 
\def \farcm{\hbox{$.\mkern-4mu^\prime$}} 
\def \fm{\hbox{$.\!\!^{\rm m}$}} 
\def \yskip{\penalty-50\vskip3pt plus3pt minus2pt} 
\def \pp{\par\yskip\noindent\hangindent 0.4in \hangafter 1} 
\def \nhat{\ifmmode {\hat{\bf n}}\else${\hat {\bf n}}$\fi}
\def \xhat{\ifmmode {\hat{\bf x}}\else${\hat {\bf x}}$\fi}
\def \yhat{\ifmmode {\hat{\bf y}}\else${\hat {\bf y}}$\fi}
\def \zhat{\ifmmode {\hat{\bf z}}\else${\hat {\bf z}}$\fi}

\title{A Determination of the Hubble Constant From Cepheid Distances
and a Model of the Local Peculiar Velocity Field}

\author{Jeffrey A. Willick\footnote{Deceased.} \  \& Puneet Batra}
\affil{Stanford University, Department of Physics, Stanford, CA 94305} 
\email{pbatra@perseus.stanford.edu}

\begin{abstract}

We present a measurement of the Hubble Constant based on Cepheid
distances to 27 galaxies within 20 Mpc.  We take the Cepheid data from
published measurements by the Hubble Telescope Key Project on the
Distance Scale ($H_0$KP).  We calibrate the Cepheid Period-Luminosity
(PL) relation with data from over 700 Cepheids in the LMC obtained by
the OGLE collaboration; we assume an LMC distance modulus of 18.50 mag
($d_{{\rm LMC}}=50.1$ kpc).  Using this PL calibration we obtain new
distances to the $H_0$KP galaxies. We correct the redshifts of these galaxies 
for peculiar velocities using two distinct velocity field
models: the phenomenological model of Tonry \etal\  and a
model based on the IRAS density field and linear gravitational
instability theory.
We combine the Cepheid distances with the corrected redshifts for the
27 galaxies to derive $H_0$, the Hubble constant.
The results are $H_0 = 85 \pm 5\,\kmsmpc$ (random error) at 95\%
confidence when the IRAS model is used, and $92 \pm 5\,\kmsmpc$ when
the phenomenological model is used.  The IRAS model is a better fit to
the data and the Hubble constant it returns is more reliable.
Systematic error stems mainly from LMC distance uncertainty which is
not directly addressed by this paper.  Our value of $H_0$ is
significantly larger than that quoted by the $H_0$KP, $H_0 = 71 \pm
6\,\kmsmpc$. Cepheid recalibration explains $\sim 30\%$
of this difference, velocity field analysis accounts for $\sim 70\%$.
We discuss in detail possible reasons for this discrepancy and future
study needed to resolve it.
\end{abstract}


\section{Introduction}

A long-standing goal of observational cosmology is the measurement of the
expansion rate of the universe, parameterized by the Hubble constant,
$H_0.$ Knowledge of $H_0$ enables us to assign galaxies absolute
distances $d$ from their redshifts $cz$,using $d=cz/H_0,$ for $z \ll 1.$ More
fundamentally, the Hubble constant measures the time since the Big
Bang, or the ``expansion age'' of the universe: $t_0=f(\omegam,\omegal)
H_0^{-1},$ where $\omegam$ and $\omegal$ are the density parameters
for mass and the cosmological constant (or ``dark energy''),
respectively.  The function $f(\omegam,\omegal)$ has well known
limiting values $f=2/3$ for $\omegam=1$ and $f=1$ for $\omegam=0,$ if
$\omegal=0.$ In the flat ($\omegam+\omegal=1$) models now favored by
CMB anisotropy measurements (e.g., Tegmark \& Zaldarriaga 2000; Lange
\etal\ 2000), $f$ is larger, at given $\omegam,$ than in the
$\omegal=0$ case.  However, unless $\omegam \simlt 0.25,$ which is
disfavored by a variety of data (Primack 2000), $f \le 1$, even in a
flat universe.  It follows that for currently acceptable values of the
density parameters, the expansion age of the universe is $\simlt
H_0^{-1}.$

By convention, extragalactic distances are measured in Mpc, redshifts
in $\kms,$ and $H_0$ in $\kmsmpc.$ From the mid-1960s through the
mid-1980s, two groups dominated the debate over $H_0.$ One, associated
mainly with Sandage and collaborators, argued that $H_0=50\,\kmsmpc$
with relatively small ($\sim 10\%$) uncertainty. A second, led by de
Vaucouleurs, advocated $H_0=100\,\kmsmpc$ with similarly small error.
The corresponding values of the expansion timescale are
$H_0^{-1}=9.8\,h^{-1}$ Gyr, where $h\equiv H_0/100\kmsmpc.$ Thus, the
large Hubble constant favored by de Vaucouleurs leads to a ``young''
($t_0 \simlt 10$ Gyr) universe, while the small $H_0$ favored by
Sandage corresponds to an ``old'' ($t_0 \simgt 15$ Gyr) universe. In
recent years, the debate has shifted, with many groups finding $H_0$
to be intermediate between the Sandage and de Vaucouleurs
values. Especially important in this regard has been the work of the
Hubble Space Telescope (HST) Key Project on the Extragalactic Distance
Scale ($H_0$KP), which finds $H_0=71 \pm 6\,\kmsmpc$ (Mould \etal\ 2000).
We discuss their methods and results further below.

An independent constraint on the age of the universe comes from the
age of the oldest stars $t_*$. This can be measured from the turnoff
point in the Hertzsprung-Russell diagrams of old globular
clusters. The best current estimates (Krauss 1999; see also Caretta
\etal\ 1999) suggest that $t_* = 12.8 \pm 1.0$ Gyr ($1\,\sigma$
error), and that $10 \le t_* \le 17$ Gyr at 95\% confidence. If one
furthermore assumes that the globular clusters did not form until
about $\Delta t \sim 1$ Gyr after the Big Bang, then the age of the
universe as indicated by the oldest stars is $t_* + \Delta t \approx
14 \pm 1$ Gyr. With the above estimates, we thus require
that $t_0$ be strictly larger than 10 Gyr, and prefer that it be
$\simgt 13$ Gyr, to ensure consistency of the Big Bang model with
stellar ages.

From this perspective, the de Vaucouleurs value of $H_0$ yields far
too small an expansion time, while the Sandage value produces one that
is comfortably large. The current modern value ($H_0$KP) gives a
marginally consistent $t_0 = 13.3 \pm 1.3$ Gyr if we assume an
$\omegam=0.3,$ $\omegal=0.7$ universe as preferred by a
variety of present data. A Hubble constant only 20\% larger, however,
would give an expansion age of 10.6 Gyr, and thus conflict with the
best estimates of the globular cluster ages.

The determination of the Hubble constant clearly remains a crucial
part of the cosmological puzzle.  Recent efforts by the $H_0$KP and other
groups have greatly reduced the allowed range for $H_0,$ but have not
unequivocally demonstrated consistency between the timescales of Big
Bang cosmology and stellar evolution. The main purpose of this paper
is to underscore the importance of ongoing work on the problem, by
presenting an alternative approach, using existing data, to measuring
$H_0.$ The outline of this paper is as follows: In \S 2 we discuss the
effects of peculiar velocities on the determination of $H_0,$ and
strategies for overcoming these effects. In \S 3 we present a
derivation of the Cepheid PL relation using the Optical Gravitational
Lensing Experiment (OGLE; Udalski \etal\ 1999a,b) database of LMC
Cepheids, and in \S 4 we apply the new PL relation to the $H_0$KP Cepheid
database to obtain distances for the $H_0$KP galaxies.  In \S 5 we
constrain the local peculiar velocity field by applying the maximum
likelihood \velmod\ method to a sample of galaxies with accurate
relative distances from surface brightness fluctuations.  In \S 6 we
apply the resulting velocity models to the $H_0$KP Cepheid galaxies, and
thereby obtain a value of $H_0.$ In \S 7 we further discuss and
summarize our results.

\section{Peculiar velocities and the strategies for measuring  $H_0$}

Measuring $H_0$ requires addressing the effects of peculiar
velocities---those deviations from the underlying Hubble flow.  The
observed redshift $cz$ of a galaxy at a distance $d$ is given by $cz = H_0 d + u,$ where $u$
is the radial component of its peculiar velocity. In the
(unfortunately) hypothetical case of pure Hubble flow ($u \equiv 0$),
a few good distance and redshift measurements of very nearby galaxies
would cleanly yield $H_0.$

Over the last two decades, however, it has become clear that galactic
peculiar velocities are both substantial ($\sim$ several hundred
$\kms$) and {\em systematic,} i.e., coherent over volumes of diameter
$\sim 10$--$20$ Mpc (see Willick 2000 for a recent review). Neglecting
peculiar velocities can produce an error $\delta H_0/H_0 \sim u/cz$
for a single galaxy, while observing $N$ galaxies does not necessarily
lead to a $\sqrt{N}$ reduction due to the coherence of the
velocity field. $H_0$ measurement errors due to uncorrected peculiar
velocities can approach $\sim 30\%$ at redshifts as large as $\sim
1500\,\kms,$ roughly the distance to the Virgo cluster, even if
moderately large ($N \sim 10$) samples are used.

At least two straightforward paradigms exist to handle the impact on
$H_0$ from peculiar velocities: In the first approach ({\em Method I}\
), one measures $H_0$ in the ``far field'' of the Hubble
flow---peculiar velocities here are washed out in comparison with the
large expansion velocities $H_0 d.$ In the second approach ({\em
Method II}\ ), one measures $H_0$ in the ``near field'' of the Hubble
flow after using an accurate model of the local pecuilar velocity field
$u(\bfr)$ to prune the observed redshifts $cz$ of their peculiar
velocity contributions.

Method I requires no knowledge of peculiar velocities, provided they
are fractionally small. At redshifts beyond $\sim 5000\,\kms,$ for
example, the fractional uncertainty in the Hubble constant, $u/cz,$
will generally be $< 10\%$ for a single galaxy. Method I does,
however, require data at distances where it is difficult or impossible
to directly employ primary distance indicators such as Cepheids.
Instead, secondary distance indicators, such as the Tully-Fisher
relation or the Surface Brightness Fluctuation method (SBF), must be
used. Not only are secondary DI's less intrinsically accurate than
Cepheids, but they also lack {\em a priori} absolute
calibrations. This motivates a first look at how sources of error in
the calibration process and distance scale relatively impact Method I
and Method II analyses.

Primarily, uncertainty in the absolute calibration of the Cepheid PL
relation, (itself due to uncertainty in the distance to the LMC),
floods the systematic uncertainty in $H_0$. As Gibson (1999) has
emphasized, values as small as $\mu_{{\rm LMC}}=18.2$ mag and as large
as $\mu_{{\rm LMC}}=18.7$ mag have appeared recently in the
literature, though $H_0$KP adopts $\mu_{{\rm LMC}}=18.50$ mag. $H_0$
estimates could be revised upwards by as much as $15 \%$ if the
smallest LMC distances prove correct, or downward by as much as $10\%$
should the largest LMC distance hold.

Also, a recalibration of the Cepheid P-L relation is desired
because the current P-L relation used by the $H_0$KP is based on only 32
Cepheids in the LMC (Ferrarese \etal\ 2000b) while the OGLE has found
many more. The $H_0$KP is also in the process of redetermining Cepheid
galaxy distances using a calibration based on the OGLE data
(W. Freedman, private communication; Madore \& Freedman 2000, in
preparation). Our own work with the OGLE data (see \S 4) indicates
that the actual Cepheid distances are systematically $\sim 5 \%$
shorter than the distances originally used by the HOKP.

\begin{deluxetable}{lccc}
\tablewidth{0pc}
\tablecolumns{4}
\tablecaption{Tonry et al. calibration of the SBF
distance scale \label{tab:zero}}
\tablehead{\colhead{} &\colhead{$\overline{ m^0_I }$} &\colhead{$\mu_{ceph}$\tablenotemark{a}} & \colhead{$\overline{ M_I }$} \\
\colhead{Galaxy} & \colhead{(mag)} & \colhead{(mag)} &\colhead{(mag)} } 
\startdata
NGC0224 &$22.67 \pm 0.06 $&$24.44 \pm 0.10$ &$-1.77 \pm 0.12$\\ 
NGC3031 &$26.21 \pm 0.25 $&$27.80 \pm 0.08$ &$-1.59 \pm 0.26$\\ 
NGC3368 &$28.34 \pm 0.21 $&$30.20 \pm 0.10$ &$-1.86 \pm 0.23$\\ 
NGC4548 &$29.68 \pm 0.54 $&$31.04 \pm 0.08$ &$-1.36 \pm 0.55$\\ 
NGC4725 &$28.87 \pm 0.34 $&$30.57 \pm 0.08$ &$-1.70 \pm 0.35$\\ 
NGC7331 &$28.85 \pm 0.16 $&$30.89 \pm 0.10$ &$-2.04 \pm 0.19$\\ 
\enddata
\tablenotetext{a}{From Ferrarese \etal\ 2000a}
\end{deluxetable}

These Cepheid uncertainties will propagate through any Method I {\em
or} II approach that uses the Cepheid distance scale. The extent of
this propagation motivates our choice of method.  Method II approaches
cut the propagation chain and use these Cepheid distances
directly. Method I approaches use this Cepheid distance scale to
calibrate a secondary distance scale, as mentioned above, which then
provides the distances used to measure $H_0$. We focus on those Method
I approaches that use Surface Brightness Fluctuations (SBF; Tonry \&
Schneider 1988) as the secondary distance indicator to highlight the
difficulties in the calibration process and offer a first motivation
toward Method II approaches.  Ferrarese \etal\ (2000a) give a more
exhaustive review of the calibration process in other secondary
distance indicators.

Table~\ref{tab:zero} shows the tabulated SBF calibration data from Tonry \etal\
(2000; hereafter TBAD00). Column 1 lists the 6 galaxy names for which
both color-adjusted SBF apparent bulge magnitudes ($\overline{ m_I }$;
Column 2) and Cepheid-based distance moduli ($\mu_{ceph}$; Column 3)
are available.

Table~\ref{tab:zero} reveals a spread in $\overline{ M_I }$ of $ \sim .8$ mag, with
half of the measured values within the estimated systematic error of
$~.16$ mag of the $H_0$KP SBF zero point, $-1.79$ mag (Ferrarese \etal\
2000a), and the Tonry \etal\ (TBAD00) zero-point of $-1.74$ mag. Both
values agree with the theoretically predicted value, $-1.81$ mag
(Worthey 1993); the $H_0$KP value leads to a distance scale that is in
good agreement with other secondary distance scales (Ferrarese \etal\
2000a).

The SBF and Cepheid Virgo distances indicate that the preferred
calibration of Tonry \etal\ (TBAD00), as well as the similar $H_0$KP
calibration (Ferrarese \etal\ 2000a), is inconsistent with our revised
Cepheid distance scale. As noted in (\S 4) the mean Cepheid distance
to the five Virgo galaxies considered in this paper is 14.7 Mpc. If
only the three galaxies within the canonical 6 degrees of the Virgo
center are considered, this rises to 15 Mpc. By comparison, the mean
SBF distance, using the Tonry \etal\ calibration above, of 27 Virgo
galaxies in the SBF sample is 16.5 Mpc. If the SBF zero point,
formerly in agreement with theory and other distance indicators, were
shifted to bring the SBF Virgo galaxies to the 15 Mpc value suggested
by the Cepheids, the value for $H_0$ derived from the analyses of
TBAD00 and Blakeslee \etal\ (1999; see \S 5) would rise by $\sim 7
\%$, while the $H_0$KP value would rise by $\sim 10 \%$, close to the
upper edge of the (Cepheid + SBF) systematic error envelope. This new
result would still depend critically on the 6 calibrating galaxies of
Table~\ref{tab:zero}. As illustrated, the inevitable errors in the
Cepheid calibration propagate directly into the secondary distance
calibration, and the derived $H_0$.

The main advantage of Method II over Method I is that no such intermediate
calibration step is required. One uses only the primary distances from
Cepheid variables, which are more reliable within $\sim 2000\,\kms$,
but still subject to uncertainties. The disadvantage of Method II is
that it requires an accurate model of the local peculiar velocity
field. Turner, Cen \& Ostriker (1992) performed theoretical analyses of the
impact a lack of an accurate velocity field would have on local
measures of $H_0$; constructing a satisfactory model has proved
difficult but possible, as we discuss below.  However, any systematic
errors in the model result in errors in the corrected redshifts, and
thus in the derived Hubble constant. The gains of removing the
systematics from secondary calibration are countered by the velocity
field systematics.

In the past decade Method I has taken precedence in Hubble constant
determinations. The $H_0$KP, in particular, has adopted the philosophy of
Method I, focusing its efforts on determining Cepheid distances to
galaxies that can serve as suitable calibrators for secondary distance
indicators.  As the foregoing discussion indicates, however, the two
approaches emphasize different systematic effects, and should each be
employed as mutual checks.  We attempt in this paper to redress the
imbalance by applying Method II. Our approach has been made possible
by the advent of data sets that allow the local velocity field to be
more accurately modeled than previously, as we discuss in detail in \S
5.
\begin{figure}[!ht]
\begin{center}
\includegraphics[scale=0.60]{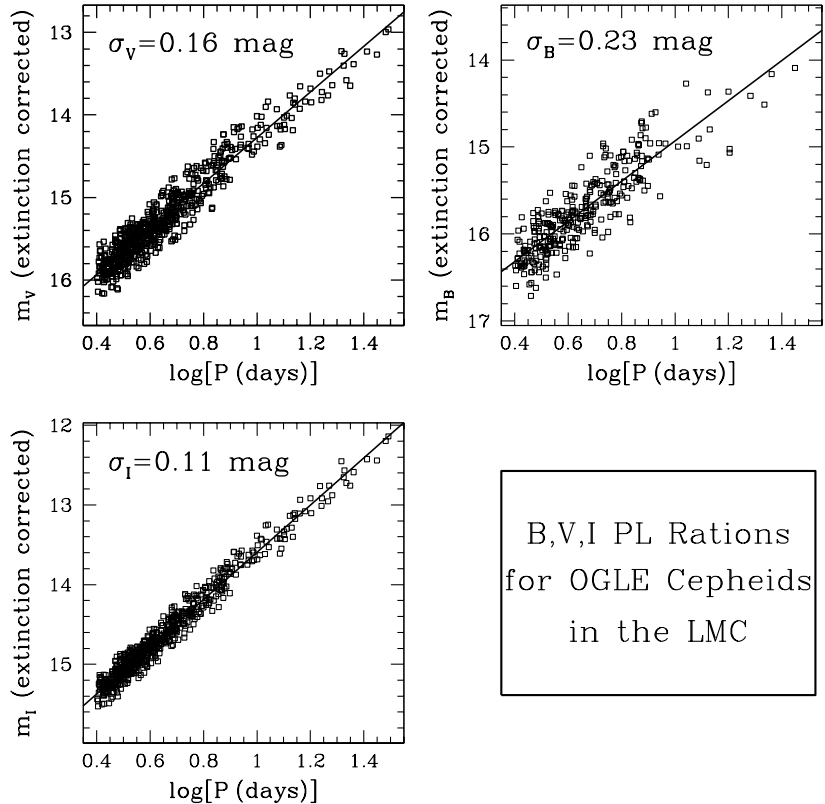}
\end{center}
\vspace{-0.5cm}
\caption{{\small B (top right), V (top left), and
I (lower left) band PL relations for Cepheid
variables found by the OGLE experiment in the Large Magellanic
Cloud. The solid lines are linear fits to the data.}}
\label{fig:f1}
\end{figure}
\section{Recalibration of the Cepheid Period-Luminosity Relation}

In \S 4 we will derive distances to the $H_0$KP sample. First, however,
we perform the recalibration of the Cepheid PL relation discussed in
\S 2, based on the OGLE data.

The OGLE project has monitored fields in the LMC nearly every clear
night for the last three years. Its primary goal was to detect
gravitational lensing events; however, in the course of this
monitoring, the OGLE discovered more than 1300 Cepheid variable stars in
the LMC.\footnote{The OGLE catalog of Cepheid variables in the LMC is
publicly available at
http://astro.princeton.edu/ogle/ogle2/var\_stars/lmc/cep/catalog.}  Of
these, a majority are fundamental mode pulsators of the sort observed
by the $H_0$KP in more distant galaxies.  We use a subsample of the OGLE
Cepheids judged by Udalski \etal\ (1999a) to be fundamental mode
pulsators, and that have periods greater than 2.5 days, to calibrate the
I and V band PL relations.  Cepheids with shorter periods may not be
fundamental mode pulsators, and moreover are not present in the $H_0$KP
data set which contains only luminous Cepheids. There are 729 such
stars in the OGLE catalog. Of these, 331 also have B band data, and we
use this subset to calibrate the B band PL relation. This latter
calibration is for reference only, however, as the $H_0$KP data consists
only of V and I band Cepheid data.

We correct the observed Cepheid mean
magnitudes for total extinction, Galactic plus LMC, as determined
by the reddenings for each field given by Udalski \etal\ (1999b).
We adopt the ratios of extinction to reddening,
$R_X \equiv A_X/E(B-V)$ (where $X=B,$$V,$$I$) 
given by Schlegel, Finkbeiner, \& Davis (1998; SFD). 
These values are given in Table~\ref{tab:OGLE}.
We then fit a linear apparent magnitude
versus log period relation in each of the three bandpasses:
\be
m_X = a_X + b_X \log P\,.
\label{eq:pl_app}
\ee
The extinction corrected magnitudes are plotted
versus $\log P$ in Figure~\ref{fig:f1}.
The solid lines show the best-fit PL relation for
each bandpass. The parameters $a_X$ and $b_X,$ along 
with related information, are given in Table~\ref{tab:OGLE}. 
Outliers were eliminated by iterating the fit several
times and excluding objects which deviated by more
than $2.5\sigma.$ 
The final fit parameters
are those obtained after this iterative procedure
converged. The number of objects used in the final
fit is given in Table~\ref{tab:OGLE} as $N_{fit},$ as compared
with the quantity $N_{tot},$ the total number of
data points in the given bandpass meeting the
initial cut on $\log P.$ The exclusion of outliers
is justified because of the likelihood of
contamination of the sample by first-overtone
Cepheids (see Udalski \etal\ 1999a for further details).

\begin{deluxetable}{ccccccc}
\tablecolumns{7}
\tablewidth{0pc}
\tablecaption{The OGLE LMC Cepheid Apparent PL Relations\label{tab:OGLE}}
\tablehead{\colhead{}&\colhead{}&\colhead{}&\colhead{}&\colhead{$\sigma$}&\colhead{}&\colhead{} \\
\colhead{Band} & \colhead{$R_X$} & \colhead{$a_X$} & \colhead{$b_X$} & \colhead{(mag)} & \colhead{$N_{fit}$} & \colhead{$N_{tot}$}}
\startdata
I &$1.96$& $16.557 \pm 0.014$ & $-2.963 \pm 0.021$ & $0.108$ & 658 & 729 \\
V &$3.24$& $17.041 \pm 0.021$ & $-2.760 \pm 0.031$ & $0.159$ & 650 & 729 \\
B &$4.32$& $17.240 \pm 0.049$ & $-2.308 \pm 0.073$ & $0.235$ & 310 & 331 \\ 
\enddata
\end{deluxetable}

\subsection{Absolute Calibration of the Cepheid PL Relation}
\newcommand{\mulmc}{\mu_{{\rm LMC}}}

The apparent magnitude-log period relations given in Table~\ref{tab:OGLE} need to
be converted into absolute PL relations in order to use them as
distance indicators for other galaxies. To do this, we must adopt a
distance modulus for the LMC, $\mulmc.$ The LMC distance has been a
contentious issue and remains, perhaps, the greatest source of
systematic uncertainty in the extraglactic distance scale (Ferrarese
\etal\ 2000b).  Recent determinations of $\mulmc$ (see Gibson 1999 for
a review) have ranged from $\mulmc=18.20 \pm 0.05$ (Popowski \& Gould
1998; Stanek, Zaritsky, \& Harris 1998; Udalski \etal\ 1998a,b) to
$\mulmc=18.55 \pm 0.05$ (Cioni \etal\ 2000; Hoyle, Shanks, \& Tanvir
2000). We will follow the $H_0$KP and adopt $\mulmc=18.50$. Although this
may be somewhat higher than the mean of recent measurements, we
believe it is the conservative choice for now. In any case, our
ultimate Hubble constant scales in a simple way with $\mulmc,$ and can
be adjusted accordingly should the LMC distance become better
determined in the future.

Using $\mulmc\equiv 18.50,$ the absolute Cepheid PL relations
follow directly from the parameters given in Table~\ref{tab:OGLE}. 
We write our
final PL relations in the form
\be
M_X = A_X + b_X ( \log P - 1 ) \,.
\label{eq:pl_abs}
\ee
We thus set the zero point at $P=10$ d,
corresponding to a ``typical'' fundamental mode Cepheid.   
Comparison of Eqs.~(\ref{eq:pl_app}) and~(\ref{eq:pl_abs}) shows
that $A_X = a_X + b_X - 18.5$ for our adopted value of $\mulmc.$
The PL slopes are of course the same.  
We thus obtain the final parameters for the Cepheid PL relations 
shown in columns (2) and (3) of Table~\ref{tab:pl}. The slope errors are
the same as those given in Table~\ref{tab:OGLE}; we do not tabulate zero point
errors because they are completely dominated by the $\sim 0.2$ mag
systematic uncertainty in the LMC distance modulus.

\begin{deluxetable}{ccccc}
\tablecolumns{5}
\tablewidth{0pc}
\tablecaption{Cepheid PL Relations ($\mu_{LMC}=18.5$  mag) \label{tab:pl}} 
\tablehead{\colhead{Band} & \colhead{$A_X$} & \colhead{$b_X$} & \colhead{$A_X$} & \colhead{$b_X$} \\
\multicolumn{1}{c}{} & \multicolumn{2}{c}{OGLE} &
\multicolumn{2}{c}{Hipparcos\tablenotemark{a}}}
\startdata
I & $-4.906 $ & $-2.963 $ & $-4.86 \pm 0.09$ & $-3.05$  \\
V & $-4.219 $ & $-2.760 $ & $-4.21 \pm 0.05$ & $-2.77$  \\
B & $-3.569 $ & $-2.308 $ & \ldots & \ldots \\ 
\enddata
\tablenotetext{a}{Lanoix \etal}
\end{deluxetable}

It is useful to compare these PL parameters to those obtained
from a completely separate sample: Cepheids from the Hipparcos
database with parallax distances. Obviously, the latter are
unaffected by the distance to the LMC, although they have
other problems, such as incompleteness and potential
systematic parallax errors. Lanoix, Paturel, \& Garnier (1999b)
have calibrated the PL relation for 174 Cepheids with Hipparcos
parallaxes and have obtained the zero point and slope given in columns
(3) and (4) of Table~\ref{tab:pl}. As can be seen, there is excellent agreement
to within the Hipparcos-based errors of the PL zero points.
The V band slopes are in similarly excellent agreement. The
I band slopes agree less well, but it is difficult
to gauge the significance of the disagreement because Lanoix \etal\ (1999b)
do not give slope errors. In any case, because $\log P = 1$ is a
typical extragalactic period, the slope difference will
not translate into a large predicted absolute magnitude
difference. Hence we can say with confidence that the
OGLE-based PL relations from the LMC are in good agreement
with the PL relations derived from Galactic Cepheids
with Hipparcos parallaxes. This argues against a large ($\simgt 0.2$ mag)
error in our adopted LMC distance modulus.

\section{Distances to Cepheid Galaxies}

We apply the OGLE PL relations given in Table~\ref{tab:OGLE} to a sample of
thirty-four galaxies for which V and I band Cepheid data are available
from the $H_0$KP team and two other groups.  It is supplemented by a
small number of Cepheids with ground-based V and I band data. All the
data were acquired from the electronic archive maintained by P.\
Lanoix at the URL {\tt
http://www-obs.univ-lyon1.fr/$\sim$planoix/ECD}; See Lanoix \etal\
(1999a) for more details.

Basic data for the thirty-four galaxies, listed in order of increasing
heliocentric redshift, are given in Table~\ref{tab:list}.  The names,
in Column 1, are those preferred by Lanoix.  Galactic longitude
($\ell$, Column 2) and latitude ($b,$ Column 3) and heliocentric
redshift ($cz_\odot,$ Column 4) were all obtained from the NASA
Extragalactic Database (NED; {\tt
http://nedwww.ipac.caltech.edu}). For reference we also list the Local
Group (LG) frame and CMB frame redshifts, $cz_{{\rm LG}}$ and
$cz_{{\rm CMB}},$ in Columns 5 and 6.  Galactic reddenings $E(B-V)$
determined by SFD (also obtained from NED) are listed in Column
7.\footnote{The SFD reddenings represent the effects of Galactic dust
only, and therefore are not suitable for correcting the Cepheid
magnitudes for extinction, which is often dominated by extinction
within the {\em host\/} galaxy.  We do not use the SFD reddenings in
any case, but present them here for completeness.} Column 8 lists the
number of Cepheids in each galaxy for which $V$ and $I$ band PL data
are available.
\begin{deluxetable}{lrrrrrcll}
\tablecolumns{9} \tablewidth{0pc} \tablecaption{List of Cepheid
Galaxies \label{tab:list}} \tabletypesize{\footnotesize}
\tablehead{\colhead{} & \multicolumn{1}{c}{$\ell$} &
\multicolumn{1}{c}{$b$} & \colhead{$cz_\odot$} & \colhead{$\;cz_{{\rm
LG}}$} & \colhead{$cz_{{\rm CMB}}$} & \colhead{$E(B-V)$} &
\colhead{} & \colhead{} \\ \colhead{Name} &
\colhead{(deg)} & \colhead{(deg)} &
\colhead{(\kms)}&\colhead{(\kms)}&\colhead{(\kms)}&\colhead{(mag)}&\colhead{$\;N_{{\rm
ceph}}$} & \colhead{Notes} } 

\startdata 
NGC 224 &$121.17$&$-21.57$&$-300$&$ -13$&$-584$&$0.062$& 37 & LG (M31) \\ IC
1613 &$129.72$&$-60.58$&$-234$&$ -65$&$-558$&$0.025$& 10 & LG\\ NGC
598 &$133.61$&$-31.33$&$-179$&$ 70$&$-459$&$0.042$& 12 & LG (M33) \\
NGC 6822 &$ 25.34$&$-18.39$&$ -57$&$ 7$&$-262$&$0.240$& 6 & LG\\ NGC
3031 &$142.09$&$ 40.90$&$ -34$&$ 126$&$ 46$&$0.080$& 25 & $H_0$KP (M81)\\
NGC 300 &$299.21$&$-79.42$&$ 144$&$ 126$&$ -89$&$0.013$& 16 & F92 \\
NGC 5457 &$102.04$&$ 59.77$&$ 241$&$ 362$&$ 360$&$0.009$& 33 & $H_0$KP
(M101) \\ SEXTANS B &$233.20$&$ 43.78$&$ 301$&$ 139$&$ 642$&$0.032$& 3
& LG \\ IC 4182 &$107.70$&$ 79.09$&$ 321$&$ 342$&$ 548$&$0.014$& 28 &
SS/KP \\ SEXTANS A &$246.15$&$ 39.88$&$ 324$&$ 117$&$ 678$&$0.044$& 7
& LG \\ NGC 3109 &$262.10$&$ 23.07$&$ 403$&$ 129$&$ 736$&$0.067$& 16 &
LG \\ NGC 5253 &$314.86$&$ 30.11$&$ 404$&$ 156$&$ 678$&$0.056$& 7 &
SS/KP \\ NGC 4258 &$138.32$&$ 68.84$&$ 448$&$ 505$&$ 651$&$0.016$& 15
& Maoz/KP \\ NGC 4548 &$285.69$&$ 76.83$&$ 486$&$ 380$&$ 805$&$0.038$&
24 & $H_0$KP (M91) \\ NGC 925 &$144.89$&$-25.17$&$ 553$&$ 781$&$
326$&$0.076$& 79 & $H_0$KP \\ NGC 2541 &$170.18$&$ 33.48$&$ 559$&$ 645$&$
694$&$0.050$& 34 & $H_0$KP \\ NGC 3198 &$171.22$&$ 54.83$&$ 663$&$ 703$&$
879$&$0.012$& 52 & $H_0$KP \\ NGC 4414 &$174.54$&$ 83.18$&$ 716$&$ 691$&$
988$&$0.019$& 11 & $H_0$KP \\ NGC 3621 &$281.21$&$ 26.10$&$ 727$&$
437$&$1059$&$0.080$& 69 & $H_0$KP \\ NGC 3627 &$241.96$&$ 64.42$&$ 727$&$
597$&$1072$&$0.032$& 36 & SS/KP \\ NGC 3319 &$175.98$&$ 59.34$&$
739$&$ 758$&$ 978$&$0.015$& 28 & $H_0$KP \\ NGC 3351 &$233.95$&$ 56.37$&$
778$&$ 640$&$1123$&$0.028$& 49 & $H_0$KP (M95) \\ NGC 7331 &$
93.72$&$-20.72$&$ 816$&$1110$&$ 492$&$0.091$& 13 & $H_0$KP \\ NGC 3368
&$234.44$&$ 57.01$&$ 897$&$ 760$&$1242$&$0.025$& 11 & T/KP \\ NGC 2090
&$239.46$&$-27.43$&$ 931$&$ 758$&$1002$&$0.040$& 34 & $H_0$KP \\ NGC 4639
&$294.30$&$ 75.99$&$1010$&$ 901$&$1328$&$0.026$& 17 & SS/KP \\ NGC
4725 &$295.08$&$ 88.36$&$1206$&$1160$&$1486$&$0.012$& 20 & $H_0$KP \\ NGC
1425 &$227.52$&$-52.60$&$1512$&$1442$&$1413$&$0.013$& 29 & $H_0$KP \\ NGC
4321 &$271.14$&$ 76.90$&$1571$&$1467$&$1893$&$0.026$& 52 & $H_0$KP (M100)
\\ NGC 1365 &$237.96$&$-54.60$&$1636$&$1546$&$1539$&$0.020$& 52 & $H_0$KP
\\ NGC 4496A &$290.56$&$ 66.33$&$1730$&$1573$&$2070$&$0.025$& 94 &
SS/KP \\ NGC 4536 &$292.95$&$ 64.73$&$1808$&$1645$&$2148$&$0.018$& 39
& SS/KP \\ NGC 1326A &$238.55$&$-56.28$&$1836$&$1750$&$1730$&$0.017$&
17 & $H_0$KP \\ NGC 4535 &$290.07$&$ 70.64$&$1961$&$1825$&$2293$&$0.019$&
50 & $H_0$KP \\ \enddata
\end{deluxetable}

Further references for each galaxy are given in column 9 of Table~\ref{tab:list}.
Seven galaxies are denoted ``LG,'' indicating that they are
members of the LG according to the compilation of Mateo (1998).\footnote{For
all seven we obtain Cepheid distances (see Table~\ref{tab:d}) smaller than
1.5 Mpc, and for four of them we obtain distances less than 0.8 Mpc. Our
distances agree with those given by Mateo (1998) to within the errors.}
The $V$ and $I$ band data compiled by Lanoix for these objects 
are ground-based. The LG galaxies are
included here for completeness, but are not used
in the $H_0$ determination presented in \S 6. Galaxies in
the LG have no leverage
on $H_0$ because their Hubble velocities are negligible
in comparison with random peculiar motions.

Of the 27 remaining objects, all of which are used in the $H_0$
analysis in \S 6, 26 have HST data. The one exception
is NGC 300, which has ground-based data from Freedman \etal\
(1992; F92). 
PL data for
this galaxy in the Lanoix database that are {\em not\/} from 
F92 are not used in our analysis. 

For the twenty-six galaxies with HST data, eighteen were  
observed originally by the $H_0$KP team. These objects
are denoted ``$H_0$KP'' in Table~\ref{tab:list}. The complete Cepheid
database for these galaxies, as well as a detailed compendium of
publications by the $H_0$KP, is given at the
Key Project website.\footnote{ The Key project web site can be found at {\tt www.ipac.caltech.edu/$H_0$KP}. The
Lanoix database includes this database as
a subset; a crosscheck confirms that the periods and magnitudes
in the two are identical.} 
A further seven galaxies were
originally observed by other HST investigators: six
by the Sandage/Saha (Saha \etal\ 1999) group, and
one by Tanvir
\etal\ (1995), denoted ``SS/KP'' and ``T/KP'' respectively, in Table~\ref{tab:list}. 
The data for all seven were subsequently
reanalyzed by the $H_0$KP team
(Gibson \etal\ 2000). We use {\em only\/} the reanalyzed
data from these seven galaxies in order to ensure uniformity with the rest of the $H_0$KP analyzed data that we use.
Finally, one galaxy, NGC 4258, is denoted ``Maoz/KP'' in Table~\ref{tab:list}. 
It is not formally part of 
the $H_0$KP sample, but was 
observed with the HST by Maoz \etal\ (1999), in collaboration
with members of the $H_0$KP team, in order 
to compare Cepheid distances with the highly
accurate maser distance of Herrnstein \etal\ (1999; we discuss
this special case further in \S 7). The Cepheid data
for NGC 4258 are as yet unpublished, but were kindly provided
to us in advance of publication by J.\ Newman. 
They are also available in Lanoix's database.

\subsection{Calculation of Distances}
\newcommand{\nceph}{N_{{\rm Ceph}}}
\newcommand{\mij}{m_{ij}}

We determine the distance to each galaxy by
minimizing a $\chi^2$ statistic that measures
deviations from the $V$ and $I$ band PL relations.
The statistic is minimized with respect to
variations in 
the galaxy distance modulus $\mu$
and the total reddening, Galactic plus internal, along the
line of sight toward each Cepheid in the galaxy.
We also assumed a single
reddening value for all stars in the galaxy. This 
led to an identical value of the distance modulus 
but a much
higher value of the PL scatter, indicating that the reddening
is variable across the face of each galaxy. 

Suppose we have PL data for $i=1,...,\nceph$ Cepheids 
in the galaxy in question, with one $V$ band and
one $I$ band mean magnitude, and one period, for each star.
Let $\mij,$ where $j=V,I,$ denote the magnitudes,
and $X_i=\log P_i,$ where $P_i$ 
is the pulsation period in days, of the $i$th star. 
The appropriate $\chi^2$ statistic is then  
\be
\chi^2 = \sum_{i=1}^{\nceph} \sum_{j=V,I} \left[ \mij - 
\left(A_j + b_j (X_j - 1) + \mu + R_j E(B-V)_{i}\right)\right]^2/{\sigma^2} \,. 
\ee
Minimization of $\chi^2$ yields the distance modulus $\mu$
and the $\nceph$ reddenings $E(B-V)_i.$ The PL parameters
for the two bandpasses are hardwired to the values given
in Table~\ref{tab:pl}. 
\begin{deluxetable}{lcrr}
\tablecolumns{4}
\tablewidth{0pc}
\tablecaption{Cepheid Galaxy Distances\label{tab:d}}
\tabletypesize{\footnotesize}
\tablehead{\colhead{}&\colhead{$\mu$}&\colhead{$d$}&\colhead{$\vev{E(B-V)}$} \\
\colhead{Name} & \colhead{(mag)} & \colhead{(Mpc)} & \colhead{(mag)}}
\startdata
  NGC224 &$24.37\pm 0.07$&$ 0.75\pm 0.03$&$0.202\pm 0.027$ \\
  IC1613 &$24.29\pm 0.14$&$ 0.72\pm 0.05$&$0.085\pm 0.052$ \\
  NGC598 &$24.47\pm 0.13$&$ 0.78\pm 0.05$&$0.223\pm 0.048$ \\
 NGC6822 &$23.27\pm 0.18$&$ 0.45\pm 0.04$&$0.337\pm 0.067$ \\
 NGC3031 &$27.66\pm 0.09$&$ 3.40\pm 0.14$&$0.138\pm 0.033$ \\
  NGC300 &$26.55\pm 0.12$&$ 2.04\pm 0.11$&$0.026\pm 0.043$ \\
 NGC5457 &$29.20\pm 0.08$&$ 6.91\pm 0.25$&$0.081\pm 0.029$ \\
    SEXB &$25.80\pm 0.26$&$ 1.45\pm 0.17$&$-.029\pm 0.095$ \\
  IC4182 &$28.27\pm 0.08$&$ 4.50\pm 0.17$&$0.016\pm 0.031$ \\
    SEXA &$25.74\pm 0.17$&$ 1.41\pm 0.11$&$0.060\pm 0.062$ \\
 NGC3109 &$25.27\pm 0.12$&$ 1.13\pm 0.06$&$0.137\pm 0.043$ \\
 NGC5253 &$27.53\pm 0.17$&$ 3.21\pm 0.25$&$0.143\pm 0.062$ \\
NGC4258  &$29.49\pm 0.12$&$ 7.90\pm 0.42$&$0.128\pm 0.043$ \\
NGC4548  &$30.94\pm 0.09$&$15.38\pm 0.64$&$0.118\pm 0.034$ \\
  NGC925 &$29.78\pm 0.05$&$ 9.04\pm 0.21$&$0.168\pm 0.019$ \\
 NGC2541 &$30.32\pm 0.08$&$11.58\pm 0.41$&$0.137\pm 0.028$ \\
NGC3198  &$30.71\pm 0.06$&$13.84\pm 0.39$&$0.101\pm 0.023$ \\
 NGC4414 &$31.17\pm 0.13$&$17.13\pm 1.06$&$0.113\pm 0.050$ \\
 NGC3621 &$29.10\pm 0.05$&$ 6.62\pm 0.16$&$0.260\pm 0.020$ \\
NGC3627  &$29.70\pm 0.07$&$ 8.73\pm 0.30$&$0.188\pm 0.027$ \\
NGC3319  &$30.70\pm 0.08$&$13.83\pm 0.53$&$0.083\pm 0.031$ \\
 NGC3351 &$29.91\pm 0.06$&$ 9.59\pm 0.28$&$0.166\pm 0.024$ \\
 NGC7331 &$30.79\pm 0.12$&$14.39\pm 0.82$&$0.196\pm 0.046$ \\
 NGC3368 &$29.94\pm 0.13$&$ 9.72\pm 0.60$&$0.155\pm 0.050$ \\
 NGC2090 &$30.30\pm 0.08$&$11.47\pm 0.40$&$0.124\pm 0.028$ \\
 NGC4639 &$31.58\pm 0.11$&$20.69\pm 1.03$&$0.117\pm 0.040$ \\
 NGC4725 &$30.37\pm 0.10$&$11.87\pm 0.54$&$0.212\pm 0.037$ \\
NGC1425  &$31.60\pm 0.08$&$20.93\pm 0.79$&$0.114\pm 0.031$ \\
 NGC4321 &$30.80\pm 0.06$&$14.48\pm 0.41$&$0.141\pm 0.023$ \\
 NGC1365 &$31.26\pm 0.06$&$17.83\pm 0.51$&$0.140\pm 0.023$ \\
NGC4496A &$30.80\pm 0.05$&$14.43\pm 0.30$&$0.107\pm 0.017$ \\
 NGC4536 &$30.77\pm 0.07$&$14.24\pm 0.47$&$0.133\pm 0.026$ \\
NGC1326A &$31.07\pm 0.11$&$16.36\pm 0.81$&$0.096\pm 0.040$ \\
 NGC4535 &$30.88\pm 0.06$&$15.02\pm 0.43$&$0.132\pm 0.023$ \\ 
\enddata
\end{deluxetable}

When we minimize this $\chi^2$ statistic for each of the 34 galaxies
listed in Table~\ref{tab:list}, we obtain the distance moduli, and
the corresponding distances in Mpc, given in
Table~{\ref{tab:d}. We discuss the calculation of uncertainties below.
We do not tabulate the reddenings for each star here,
although this information can be made available electronically
to interested readers. We do, however, list the mean
reddenings, $\vev{E(B-V)},$ for each galaxy in Table~\ref{tab:d}. We note
that, to within the errors, these reddenings are consistent with
being larger than the SFD Galactic reddenings given in Table~\ref{tab:list}.
When the reddening errors are sufficiently small for an
estimate to be made---typically, when $\nceph \simgt 30$---we find
that the mean {\em internal\/} reddening,
$\vev{E(B-V)_{int}} = \vev{E(B-V)} - E(B-V)_{SFD98}$, is
of order $0.1$ mag.

Once a galaxy distance modulus and the $\nceph$ reddenings are determined,
one may calculate $V$ and $I$ band absolute magnitudes for each
star as follows:
\be
M_{i,j} = \mij - R_j E(B-V)_{i} - \mu\,,
\label{eq:absmag}
\ee
where $i=1,...,\nceph$ and $j=V,I.$ The PL relation for the
entire sample may then be exhibited as a plot of $M_{i,j}$ versus
$\log P_i$ for all stars in all galaxies,  as shown
in Figure~\ref{fig:f2}. A total of 1021 stars is plotted
in the Figure. The OGLE PL relations given in Table~\ref{tab:pl} are
plotted through the points as dotted lines. Note that the
$V$ band PL relation has smaller scatter than the $I$ band
relation. In fact, the scatter visible in the plots is
smaller than the true scatter, because the fits have one
degree of freedom, the reddening, for each star (plus one for each galaxy).

\begin{figure}[!ht]
\begin{center}
\includegraphics[scale=0.650]{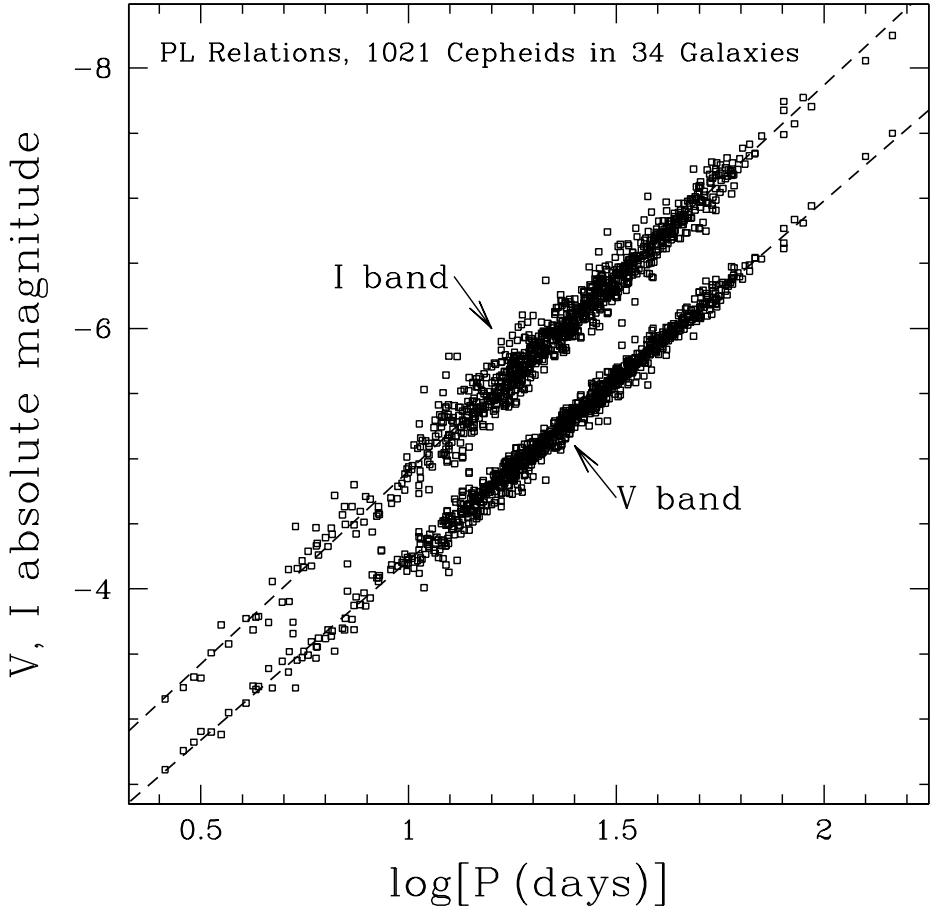}
\end{center}
\vspace{-0.3cm}
\caption{{\small $V-$ and $I-$ band PL relations for the
1021 Cepheids in the 34-galaxy sample described in Table~\ref{tab:list}.
The periods are the measured ones, while the absolute
magnitudes are calculated using Eq.~(\ref{eq:absmag}).
The dotted lines show the OGLE PL calibrations
given in Table~\ref{tab:pl}.}}
\label{fig:f2}
\end{figure}

\newpage

\newcommand{\sigceph}{\sigma_{{\rm Ceph}}}
When these degrees of freedom are properly taken into
account, the PL scatters for the 1021 Cepheids are 
\be\begin{array}{rcll}
\sigma_V & = & 0.111 & {\rm mag} \\ 
\sigma_I & = & 0.165 & {\rm mag}\,. \\ 
\end{array}
\ee
These scatters are similar to those derived
from the LMC fit (Table~\ref{tab:OGLE}), although there it was
the $I$ band scatter that was smaller. This most
likely reflects the fact that in the LMC fit the
reddenings were given a priori, and reddening errors
increase the $V$ band PL residuals more than $I$ band residuals.
Taken together, the LMC and 34-galaxy sample results suggest
that the $V$ and $I$ band PL scatters are about the same
in the absence of extinction, and that
$\sigma_V \approx \sigma_I \approx \sigceph = 0.15$ mag.
We assume that this approximation holds for
the remainder of the paper. Our method for calculating
random distance errors for the Cepheid galaxies is given
in Appendix A.

\section{Fitting Velocity Field Models} 

In order to use the Cepheid galaxies to estimate
the Hubble constant, we must, as explained in 
\S~2, use an accurate model of the local velocity field.
We employ two quite different models in this paper:
\begin{enumerate}
\item An IRAS model, in which peculiar velocities are 
based on the distribution of galaxies in the nearby
universe ($cz \simlt 10,000\,\kms$) as determined by
the 1.2 Jy IRAS redshift survey (Strauss \etal\ 1992; Fisher \etal\ 1995). 
The velocity field is then a function
of the parameter $\beta\equiv\omegam^{0.6}/b,$ where $b$ is
the biasing factor for IRAS galaxies. Before applying
the model the IRAS model to the Cepheids, then, we
must first determine the appropriate value(s) of $\beta$
to use. 
\item A phenomenological model, 
in which the local velocity field
is dominated by a few simple components. We adopt a
model of the form recently used by TBAD00,
which includes a dipole, a quadrupole, and two ``attractors,''
one centered on the Virgo Cluster and one on the Great Attractor (GA).
This model has a large number of free parameters, 
whose values must be determined before the model can
be applied to the Cepheids.
\end{enumerate}

Each of the two models has strong and weak points. The IRAS
model is more realistic and is better motivated physically. {\em All\/}
mass fluctuations in the local universe, not only prominent attractors,
affect the velocity field, as they must if structure grows
from gravitational instability. However, the IRAS model suffers
from undercounting early-type galaxies in clusters, and thus
may well underestimate the importance of massive concentrations
such as Virgo and the GA. In contrast, the Tonry (TBAD00) model
allows one to ascribe as much influence on the velocity
field to Virgo and the
GA as the data warrant. However, because it includes
{\em only\/} these attractors, the Tonry model may attribute greater or
lesser importance to these than they would have
in reality to compensate for missing 
mass concentrations and voids. 

The fact that each model is imperfect suggests that the prudent
approach is to try both. In what follows, we 
first describe (\S 5.1) our method for fitting velocity models.
We next describe (\S 5.2) the $281$
galaxy Surface Brightness
Fluctuation (SBF) data set to which we fit each of the two models
and comment on a key difference between our treatment of
this data set and previous treatments.
After that we present (\S 5.3) the constraints that the SBF data set
allows us to place on the IRAS model (i.e., the allowed
values of $\beta$ and several ancillary parameters
to be described), and (\S 5.4) the best-fitting 
parameters of the phenomenological Tonry model.

\subsection{The \velmod\ Method}
\label{sec:velmod}

For both optimizing our peculiar velocity models with the SBF data,
and for determining $H_0$ with the Cepheid data (in \S 6), we use the
\velmod\ maximum likelihood approach.  The method was described in
detail in two papers (Willick \etal\ 1997b, hereafter WSDK; Willick \&
Strauss 1998, hereafter WS), so we limit ourselves to a brief overview
here. When we fit a peculiar velocity model to redshift-distance data,
two sources of variance enter in: distance measurement error and
small-scale velocity noise.  The latter is the part of the the
peculiar velocity field that cannot be predicted by any model. It
results from close gravitational encounters with other galaxies in
groups, rather than from the systematic pull of the large-scale
gravity field.  When velocity models are fit to relatively distant
($\simgt 30\hmpc$) galaxies, the distance errors ($\simlt 10\%$ for
SBF or Cepheid distances, $\sim 20\%$ for Tully-Fisher distances)
dominate.  However, at the much smaller distances of interest here
($\simlt 20\hmpc$), the velocity noise is equal to or larger than
distance errors. Thus, it is essential that we calibrate it properly.

\velmod\ was designed with precisely this aim. It
is based on the explicit expression for the probability
that a galaxy along a given line of sight has a measured
distance $d$ and redshift $cz:$
\be
P(\ln d,cz; \bfp) \propto \zinfint r^2 n(r) P(\ln d|r) P(cz|r)\,dr \,,
\label{eq:pdcz}
\ee
where $r$ is the {\em true\/} (as opposed to measured)
distance along the line of sight, and $\bfp$ is
a vector of parameters that determine the
model peculiar velocity field.
Here and in what
follows, we assume that
$r$ and $d$ are measured in units of $\kms,$ i.e.,
we take $H_0 \equiv 1.$ We will drop this assumption
only in the last step of the analysis, when we
apply \velmod\ to the Cepheid sample to determine
$H_0.$  In Eq.~(\ref{eq:pdcz}), 
$n(r)$ is the number density of galaxies at distance
$r$ along the line of sight, 
\be
P(\ln d|r) = \frac{1}{\sqrt{2\pi}\Delta} 
\exp\left\{-\frac{\left[\ln(d/r)\right]^2}
{2\Delta^2}\right\} \,,
\label{eq:pdr}
\ee
where $\Delta = 0.46\delta\mu$ is the fractional distance error,
$\delta\mu$ is the distance modulus error, and
\be
P(cz|r;\bfp) = \frac{1}{\sqrt{2\pi}\sigma_v(r)} 
\exp\left\{-\frac{\left(cz
-[r+u(r)]\right)^2}{2\sigma_v(r)^2}\right\} \,,
\label{eq:pczr}
\ee
where $\sigma_v(r)$ is the velocity noise 
and $u(r)$ the radial component of the predicted
velocity field. 
Note that we allow $\sigma_v$ to vary with
position; in practice we can also allow it (in the
case of the IRAS model) to vary with local number
density. Of course, $u(r),$ $\sigma_v(r),$ and $n(r)$
depend on the parameter
vector $\bfp.$

Eq.~(\ref{eq:pdcz}), the joint probability distribution
of distance and redshift, is not the optimal quantity on
which to base likelihood-maximization because it is quite
sensitive to the density as well as the velocity model. 
It is more suitable to use the conditional probability,
\be
P(\ln d|cz;\bfp) = \frac{P(\ln d,cz)}{\zinfint P(\ln d,cz)\,d(\ln d)}
= \frac{\zinfint r^2 n(r) P(\ln d|r) P(cz|r)\,dr}
{\zinfint r^2 n(r) P(cz|r) \,dr} \,,
\label{eq:pdgcz}
\ee
which is less sensitive to the density model 
because of the presence of $n(r)$ in both
numerator and denominator. The conditional probability
that the $i=1,...,N$ sample galaxies have observed
distances $d_i$ given that their redshifts are $cz_i$
is then 
\be
P({\rm data};\bfp) = \prod_{i=1}^N P(\ln d_i|cz_i; \bfp)\,.
\label{eq:probdata}
\ee
The essential step in \velmod\ is maximizing the above sample
probability with respect to the model parameter
vector $\bfp.$ In practice, this is done by minimizing
the statistic
\be
{\cal L} = -2\ln P({\rm data};\bfp) \,.
\label{eq:defcall}
\ee
WSDK showed using simulated data
sets that minimizing ${\cal L,}$ as defined by
Eqs.~(\ref{eq:pdgcz}) through~(\ref{eq:defcall}),
recovers unbiased values of velocity model parameters.

\subsection{The SBF Sample}

The best current data set to use for constraining
the local peculiar velocity field is the SBF sample
of Tonry and collaborators (Tonry \etal\ 1997,
2001). The full sample comprises
$\sim 300$ early-type (mainly E and S0) galaxies out to
$\sim 4000\,\kms,$ although most
are within $\sim 3000\,\kms.$ For our fitting, we use
a subset of the sample consisting of 281 galaxies that
are not in the LG, have $(V-I)$ colors $> 0.9,$ and
are consistent with our velocity models at the $\le 3\,\sigma$ level.
The SBF distances are accurate in
most cases to $\simlt 10\%.$ This accuracy is considerably
better than what is available for Tully-Fisher samples;
moreover, the SBF method yields a reliable distance {\em error\/}
estimate for each galaxy, whereas for Tully-Fisher one generally
has only a global scatter which is itself somewhat uncertain.
Having a good distance error estimate is crucial if the
velocity noise information is to be properly incorporated
into the \velmod\ procedure.

TBAD00 and Blakeslee \etal\ (1999) have already used the SBF sample to
fit local velocity models. The former study fit the phenomenological
model mentioned above, and the latter an IRAS model, to the SBF
distances. Neither study, however, used the \velmod\ method per
se. Tonry \etal\ used a related approach, one that accounted for
small-scale velocity dispersion but not for the role of the volume
element (the $r^2$ term in \S 5.1) or of density variations (the
$n(r)$ term in \S 5.1).  Blakeslee \etal\ employed the same method
that Davis, Nusser, \& Willick (1996) used to fit a Tully-Fisher
sample at larger distances; as this approach does not fully account
for the effects of velocity noise, it is less suitable for the nearby
flow field analysis that is of greatest importance here.  In what
follows, we borrow from both the TBAD00 and the Blakelee \etal\ (1999)
studies, in that we employ similar models of the velocity field, but
we use the \velmod\ method in order to treat effects neglected in
those papers.

Unlike TBAD00 and Blakeslee \etal\ (1999), we do not assume an
absolute calibration of the SBF relation.  That is, we use the SBF
data only as indicators of distances in $\kms,$ not in
Mpc. Specifically, Tonry \etal\ and Blakeslee \etal\ took the SBF
relation to be \be \overline{ M_I } = -1.74 - 4.5 (V-I) \,,
\label{eq:sbf}
\ee   
where $\overline{M_I}$ is the mean fluctuation absolute magnitude
of a galaxy of a given $(V-I)$ color. They then assign
an absolute distance $d_{abs} = 10^{\left[0.2(\overline{m_I}-
\overline{M_I})-5\right]}$ Mpc to a galaxy with measured
apparent fluctuation magnitude $\overline{m_I}.$ Consequently,
when they fit their velocity models, one of the unknown
free parameters is necessarily the Hubble constant\footnote{Indeed,
TBAD00 reported $H_0=77 \pm 4\,\kmsmpc$ and Blakelee
\etal\ (1999) reported $H_0=74 \pm 4\,\kmsmpc$ (random errors) in addition
to their constraints on the velocity field.}. 
In contrast, we write the SBF relation
\be
\overline{M_I} = A - 4.5 (V-I) \,,
\label{eq:sbf_us}
\ee
where $A$ is a free parameter in our maximum likelihood analysis,
and then compute SBF distances in $\kms$ according to
$d=10^{\left[0.2(\overline{m_I}-
\overline{M_I})\right]}\,;$ it is this value of $d$ that
enters into calculation of the probability, Eq.~(\ref{eq:pdgcz}). 
Because we never convert SBF distances to Mpc, {\em our analysis
of the velocity field using the SBF data does not yield
a value for, and is completely unrelated to, the value of
the Hubble constant.} That is, no assumptions concerning the distance
scale go into our velocity field analysis; it is only later,
when we apply the \velmod\ method to the Cepheid galaxies,
that absolute distances enter our analysis. Therefore, it
is only the Cepheid distances themselves that directly affect
the value of the Hubble constant. 
(We emphasize, however, that we adopt the Tonry \etal\ SBF slope
($4.5$) as well as their reported distance errors in our analysis.)

It should be noted that our use of a relative rather than an
absolute zero point, while conceptually important, 
is largely a technical distinction.
The values of $H_0$ obtained by Tonry \etal\ and
Blakeslee \etal\ are entirely dependent on their specific choice
of SBF zero point in Eq.~(\ref{eq:sbf}), and this value is
not very well determined (see Tonry \etal\ 1997 for an
in-depth discussion). A different choice of zero point would
have led Tonry \etal\ and Blakelee \etal\ to find a different
$H_0,$ but not a different velocity field.
Our approach simply makes this explicit
by removing absolute distances altogether from the
problem of determining
the velocity field.

\begin{figure}[!ht]
\begin{center}
\includegraphics[scale=0.650]{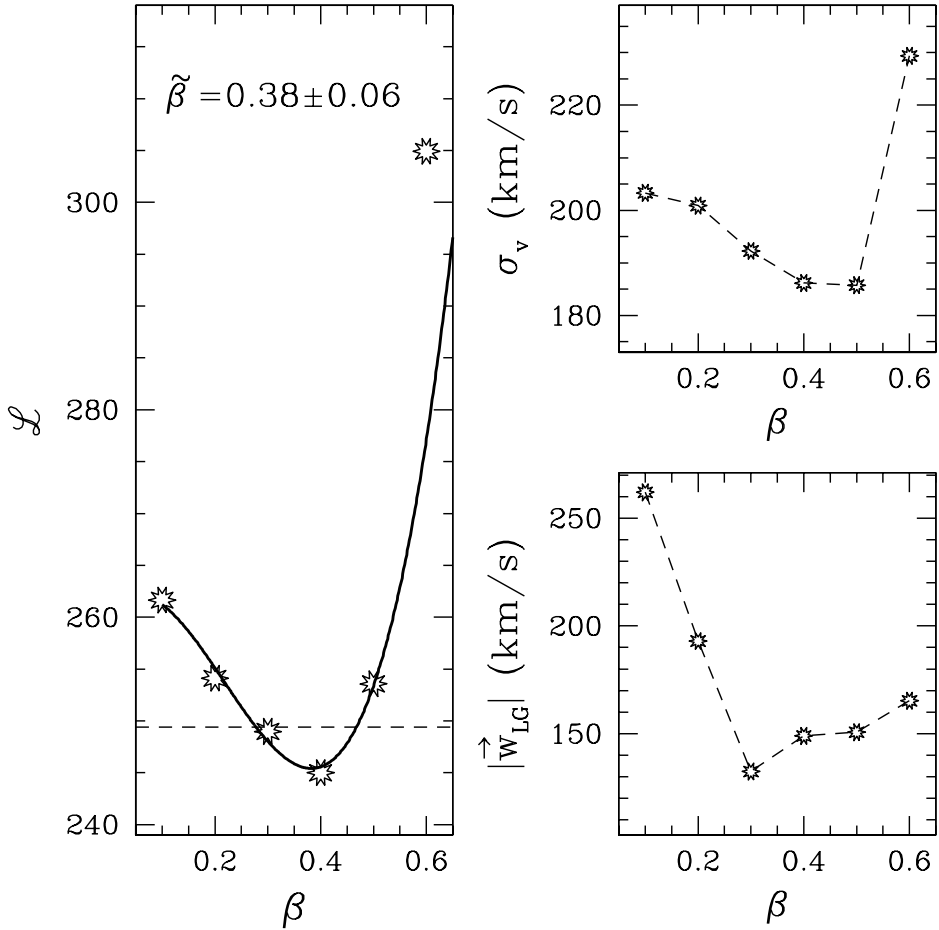}
\end{center}
\vspace{-0.3cm}
\caption{{\small {\it Left}: The points show the
\velmod\ likelihood statistic
${\cal L}$ versus $\beta$ for the SBF sample and the linear,
$300\,\kms$-smoothed IRAS velocity field (see text for details).
The solid line shows a cubic fit to the points for $0.1 \le \beta \le 0.5.$
The minimum of the curve yields the maximum likelihood value
of $\beta$ and its uncertainty, as indicated on the figure.
The horizontal dashed line shows the $2\sigma$ confidence
limits on $\beta.$ {\it Top right}: the maximum-likelihood value
of the velocity noise $\sigma_v$ versus $\beta.$ {\it Bottom right}:
Maximum-likelihood value of the Local Group velocity amplitude,
$|\bfw_{LG}|,$ relative to the IRAS predictions.}} 
\label{fig:f3}
\end{figure}

\subsection{The IRAS Model}

There are any number of ``IRAS models'' one can apply because,
aside from the obvious question of the value of $\beta,$
there are the issues of smoothing scale, filtering, nonlinear
corrections to the velocity-density relation, etc. The range of
possibilities was discussed by WSDK and WS, to which readers
are referred for a detailed discussion of the relative merits of each.
Tests in those papers with Tully-Fisher data, and with
the SBF data (Willick, Narayanan, Strauss, \& Blakeslee, in preparation;
hereafter WNSB),
have demonstrated clearly that the smallest possible Gaussian
smoothing scale for the IRAS data, $300\,\kms,$ yields the
most accurate velocity field (see Berlind, Narayanan, \& Weinberg 2000
for a theoretical justification of this fact).  A second
outcome of such tests is that the best fits to both the 
Tully-Fisher and SBF data are obtained when the linear theory
velocity-density relation,
\be
\bfv(\bfr) = \frac{\beta}{4\pi} \int d^3\bfr'\,\frac{\delta_g(\bfr')
(\bfr' - \bfr)}{\left|\bfr' - \bfr\right|^3} \,,
\label{eq:vpdelta}
\ee
is used, where $\delta_g$ is the density contrast of IRAS
galaxies smoothed on a 300 \kms\ scale. Proposed modifications to
Eq.~(\ref{eq:vpdelta}) to account for nonlinear dynamics,
as well as hypothesized nonlinear biasing relations, have not yet
been shown to improve the fit, although they typically increase the best
value of $\beta$ by $\sim 10\%.$ The reasons for this are
unclear at present, but will be discussed in depth
by WNSB. For now, we use
the linear theory, $300\,\kms$ smoothed IRAS velocity
field based on Eq.~(\ref{eq:vpdelta}).\footnote{In the language
of WSDK and WS, we note that our adopted model also employs 
Wiener filtering and Method IV.}

Figure~\ref{fig:f3} shows the main results of applying 
\velmod\ to the SBF data set using the linear IRAS velocity
field. The ${\cal L}$ versus $\beta$ plot shows a strong minimum
(likelihood maximum) near $\beta=0.4;$ the 
solid curve is a cubic fit to the likelihood points,
and its minimum determines the maximum likelihood
value of $\beta$ and its $1\,\sigma$ uncertainty:
$\beta=0.38 \pm 0.06.$ 
The $\beta=0.3$ likelihood lies within $4$ units of ${\cal L}$
from the minimum and is thus
acceptable at the $2\,\sigma$ level,
while the $\beta=0.2$ and $\beta=0.5$ models are about
$3\,\sigma$ away from the maximum likelihood value.
Note that $\beta > 0.5$ is ruled out with very high confidence. 
Indeed, we do not plot results for $\beta > 0.6$ since
the likelihood is so poor. These results suggest that we
apply the $\beta=0.2,0.3,0.4,$ and $0.5$ IRAS models to
the Cepheid galaxies when we determine $H_0$ below, giving
the most weight to the $\beta=0.3$ and $\beta=0.4$ results.

As is standard for IRAS \velmod\ 
(WSKD), we also allow for a random motion of the LG $\bfw_{{\rm LG}}$
with respect to the IRAS prediction, which is treated as a free parameter.
As found by WSKD and WS, this velocity is generally small
$\simlt 150\,\kms$ for $\beta$ near its optimum value.
This result is confirmed here in the lower right panel of
Figure~\ref{fig:f3}. The upper right panel shows
the variation of the final free parameter, the velocity
noise $\sigma_v,$ with $\beta.$ It, too, generally minimizes
near the maximum likelihood $\beta,$ and this occurs
here as well. However, note that $\sigma_v$ has a considerably
larger value, $\sim 185\,\kms,$ for the best fit model than
was found in the Tully-Fisher \velmod\ fits of WS and WSKD,
where $\sigma_v \approx 130$--$150\,\kms$ was a more typical
value. It is probable that this reflects a real difference
between the small-scale velocity noise for late-type
(Tully-Fisher) and early-type (SBF) galaxies; we will
address this issue in WNSB.
When we apply the IRAS (and Tonry) models to the Cepheid galaxies in \S 6,
we will use values of $\sigma_v$ typical of the Tully-Fisher
spirals, rather than the large $\sigma_v$ found here.

\begin{figure}[!ht]
\begin{center}
\includegraphics[scale=0.750]{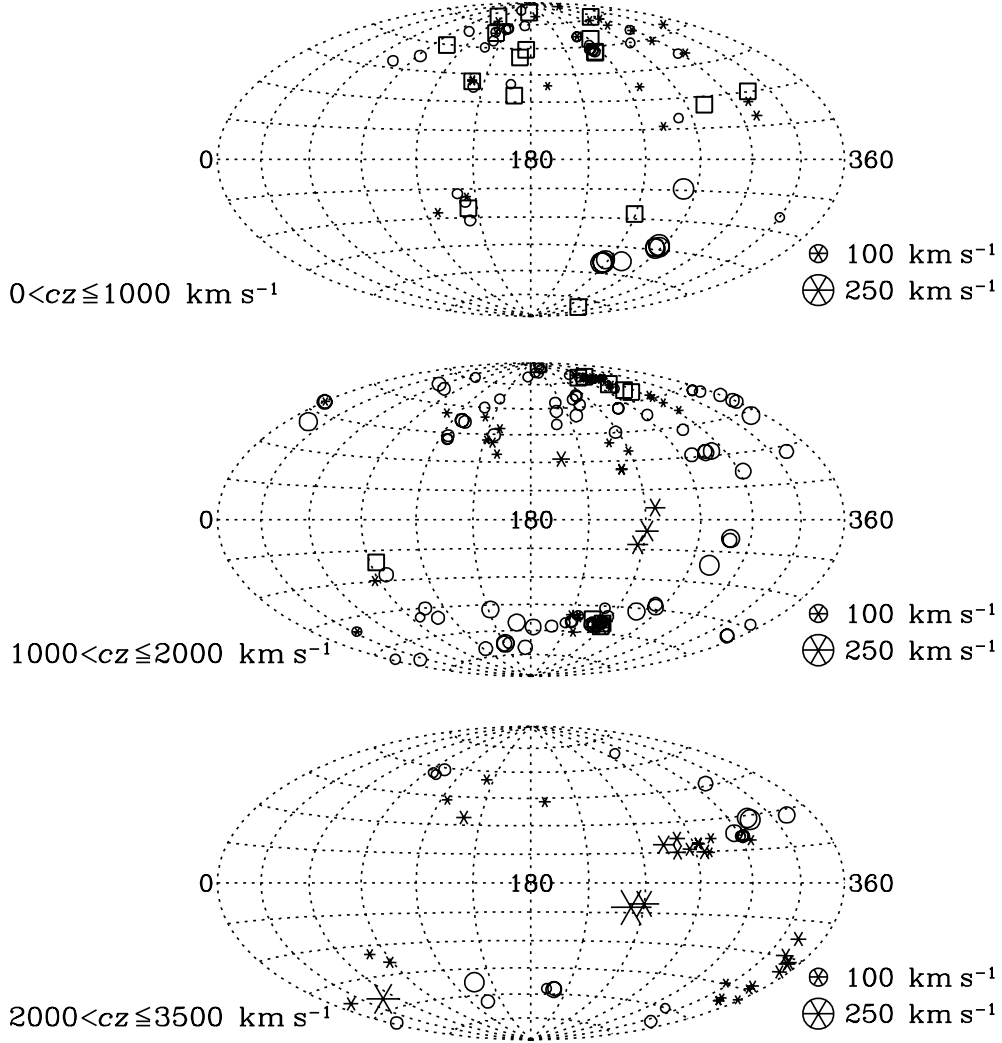}
\end{center}
\vspace{-0.3cm}
\caption{{\small Smoothed peculiar velocity residuals of the 
SBF galaxies relative to the IRAS-predicted velocities,
for $\beta=0.4.$ Stars represent objects
with positive radial peculiar velocity relative to the IRAS model, 
and open circles objects with negative peculiar velocity
relative to IRAS. The open squares show the positions
of the 27 Cepheid galaxies used to determine $H_0.$}}
\label{fig:f4}
\end{figure}

As discussed by WSDK and WS, a good method for visually assessing the
quality of the IRAS velocity model is to compute smoothed velocity
residuals for the sample (see, e.g., Eq.~(9) of WS for details on how
such residuals are calculated). Sky plots of such residuals are
presented in Figure~\ref{fig:f4} for the $\beta=0.4$ model. The
Gaussian smoothing scale employed rises from $250\,\kms$ at $cz \simlt
500\,\kms$ to $700\,\kms$ at $3500\,\kms$; thus, the smoothed
residuals will exhibit coherence over $\sim 30\degs$ scales. Coherence
on larger scales than this signifies a failure of the model.

The residuals in Figure~\ref{fig:f4} are essentially
incoherent on large scales, as indicated by the fact that
both starred (outflowing) and open (inflowing) symbols are
well mixed throughout the plots.
Furthermore, the amplitude of these velocity residuals is almost
everywhere $\simlt 100\,\kms,$ and is in many places virtually
zero (indicated by the smallest points on the figure).
These aspects of Figure~\ref{fig:f4} indicate that the IRAS model
provides a good fit to the SBF data. The open squares in
Figure~\ref{fig:f4} indicate the positions of the Cepheid galaxies
to which the IRAS and Tonry models will be applied
in \S 5. The squares are seen to lie in regions
well sampled by the SBF data, so that the velocity models
are well-constrained where the Cepheid galaxies are found.

There is one region, however, where a conspicuous failure
of the IRAS model seems to occur: in the lower right
quadrant of the $cz \le 1000\,\kms$ map. A group of
six SBF galaxies appear to have a coherent flow at
$\sim 150\,\kms$ relative to the IRAS model there.
However, this pattern does not continue into the
next redshift interval, where there are more Cepheid galaxies
and where the leverage on the $H_0$ measurement is
greater. Consequently, we do not attempt to correct
for this apparent failure of the IRAS model (the
Tonry model shows the same discrepancy). However, this
anomaly remains unexplained and deserves further
attention.

\subsection{The Tonry Model}

Our phenomenological model has the same functional form as
the TBAD00 model. 
However, we have implemented the model
in a system of units in which distance is measured in \kms,
and used the \velmod\ formalism, which differs in a number
of ways (\S\ref{sec:velmod}, Batra \& Willick 2000 ) from that of TBAD00. 
To maximize independence from the IRAS
results, we assume $n(r)=\,$constant in
Eqs.~(\ref{eq:pdcz}) and (\ref{eq:pdgcz}) when we
implement the Tonry model. 

An important distinction between the Tonry model and the IRAS models
is that the former is carried out entirely in the CMB reference frame,
i.e., the redshifts used are $cz_{{\rm CMB}},$ whereas for the IRAS
models $cz_{{\rm LG}}$ is used. The frame of reference is a highly
nontrivial issue for local velocity field fits, as redshifts can
differ by up to $600\,\kms$ (Table~\ref{tab:list}, Courteau \& van den Bergh
1999). It is thus important to demonstrate that comparable results can
be obtained regardless of frame.\footnote{One can also implement the
Tonry model in the LG frame. We have done this, and the value of $H_0$
we obtain is unchanged.}

The Tonry model assumes the local peculiar velocity field 
to be dominated by two spherically symmetric attractors, one
centered on the Virgo Cluster and one on the Great Attractor.
Each attractor is taken to have a mean interior overdensity given by
\begin{equation}
\delta(r) = \frac{\delta_0 e^{ -r/R_{cut}}}{1 - \gamma/3} \left(\frac{r}{R_{c}}\right)^{-3} \left[\left(1 + (r/R_c)^3\right)^{1- \gamma/3} - 1 \right]
\end{equation}
where $r$ is the distance from the center of the attractor. 
Note that the attractors have both
a ``cutoff'' radius $R_{cut}$ and a ``core'' radius $R_c.$
Each attractor produces an infall velocity given by the 
Yahil's (1985) expression,
\begin{equation}
v_{in}(r)=\frac{1}{3} \omegam^{0.6} r\, \delta(r) 
\left[1 + \delta(r)\right]^{-1/4}
\end{equation}
The value of $\omegam$ is not well constrained by the
fit, being highly covariant with $\delta_0$ (TBAD00), 
and it thus suffices to fix it at an arbitrary value,
which we take to be $0.2.$  

To roughly account for the effects of mass inhomogeneities
other than the attractors, the model also includes 
velocity dipole and quadrupole fields; the latter is
exponentially truncated and centered
on the LG. These fields are added vectorially to the
infall velocities produced by the attractors. 

We fit the Tonry model to the SBF data set, allowing some,
but not all, of the model parameters to vary. As discussed
in detail by TBAD00, there is significant covariance
among the parameters, so to simplify the fit
we take the attractor core radii $R_c$ and power-law
exponents $\gamma$ to have the TBAD00 values. The model
also contains the velocity dispersions of the
tracer galaxies as a function of position (i.e., the
quantity $\sigma_v(r)$ in Eq.~(\ref{eq:pczr})). TBAD00
took the background dispersion to be $187\,\kms,$  and
then added this value in quadrature with an additional
dispersion, $\sigma_v^{core},$ within a distance $R_c$ 
of three clusters: Virgo, the Great Attractor, and Fornax.
For these dispersions, too, we adopted the TBAD00 values.
(Note, however, that Fornax does not contribute to the
overall velocity field.)

Table~\ref{tab:ton} presents our best fit values
for the parameters we allowed to vary, as well as
those held fixed at the TBAD00 values.
Those parameters which denote positions in space
are given in Supergalactic coordinates, in $\kms$ units. For this table, the dipole components, in $\kms$, are $(-70, 230, -40)$; the quadrupole matrix is
\begin{equation}
\nonumber
\left( \begin{array}{ccc}
1.79&2.17&-6.62 \\
2.17&-10.5&-3.99 \\
-6.62 & -3.99&8.71 \\
\end{array} \right) .
\end{equation}

\begin{deluxetable}{lcccccc}
\tablecolumns{7}
\tablewidth{0pc}
\tablecaption{Best-fit parameters of Tonry Model\label{tab:ton}}
\tablehead{
 \colhead{} & \colhead{(x,y,z)}  & \colhead{$\sigma_{v}$\tablenotemark{a}} & \colhead{} &\colhead{$R_{cut}$} & \colhead{} & \colhead{$R_c$\tablenotemark{a}}\\
\colhead{Attractor} & \colhead{(\kms)} &\colhead{(\kms)}&\colhead{$\delta_0 $}&\colhead{(\kms)} & \colhead{$\gamma$\tablenotemark{a}} &\colhead{(\kms)} }
\startdata
Virgo &$(-250, 1300, -140)$& $650$ & 54.7 & 968 & 1.5 & 157 \\ 
GA &$(-2950, 1170, -1370)$& $500$ &  179.4 & 4010 & 2.0 & 157  \\ 
Fornax &$(-150, 1180, -1050)$& $ 235$ & N/A &N/A&N/A& 157\\ 
\enddata
\tablenotetext{a}{Parameter held fixed at the TBAD00 value.} 
\end{deluxetable}

\subsubsection{Comparison of Velocity Models}

Figure~\ref{fig:f5} 
provides a comparison of the IRAS and Tonry models along the
lines of sight toward four SBF galaxies that lie in regions
that are also important for the Cepheid analysis of \S 6.
We plot both our own fit of the Tonry model to the SBF
data (solid line) and that of TBAD00 (dotted line).
Peculiar velocity in the LG frame is plotted as a function
of Hubble-flow distance. The IRAS velocity
field generally exhibits more ``features,'' i.e., it is
not as smooth as the Tonry models. This results from the
fact that the IRAS peculiar velocity field is produced by
all mass fluctuations, while the phenomenological model
assumes only the existence of two attractors. Another
important difference is in the different strength of the
Virgo (NGC 4476) and Great (NGC 4616) Attractors. The IRAS
velocity field exhibits relatively weak gradients 
($|u'(r)| \simlt 0.5$ near Virgo, $|u'(r)| \simlt 0.25$ near
the Great Attractor), while the Tonry model has large
gradients ($|u'(r)| \simgt 1$) in these regions. The
greater influence of the attractors 
arises because, in the Tonry model, they must account for all features
of the velocity field, some of which are in reality 
due to other mass fluctuations. 
(On the other hand, the mild gradients in
the IRAS $u(r)$ near Virgo may be due in part to
the undercounting of cluster galaxies by IRAS.)   
As we shall see \S 6, the Tonry
model does not fit the Cepheid data as well as the IRAS 
model; as a result, we shall, in the end, adopt the value of $H_0$ derived
from the IRAS model. 

\begin{figure}[!ht]
\begin{center}
\includegraphics[width=5in]{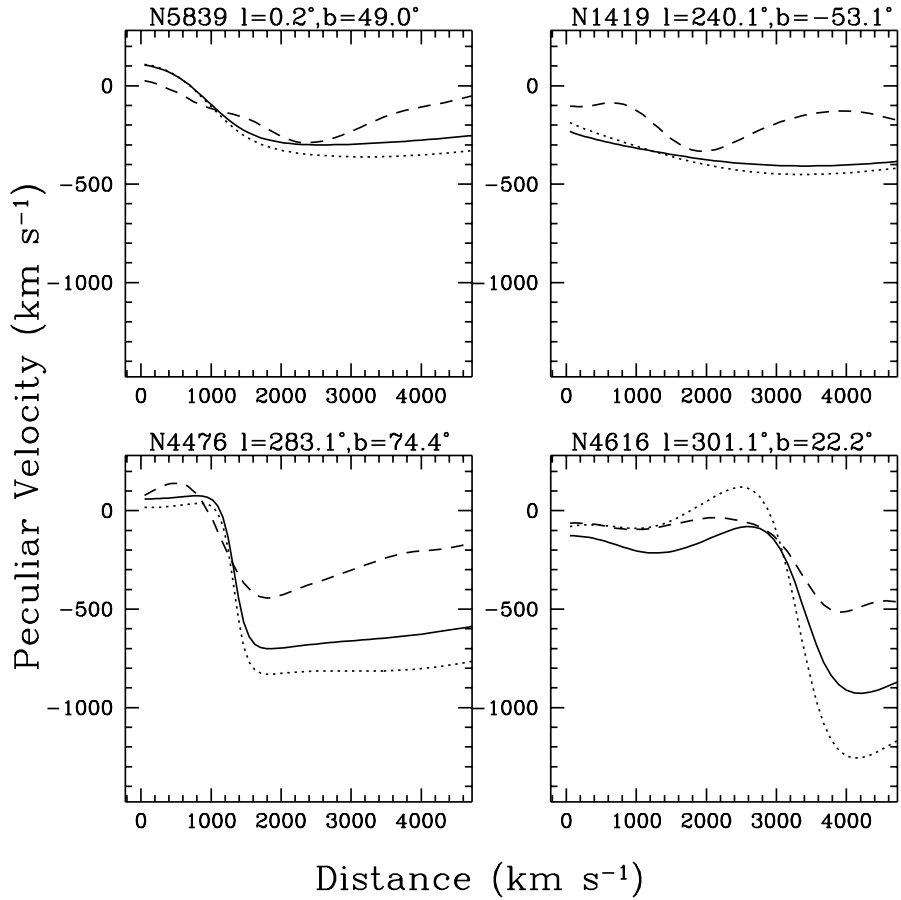}
\end{center}
\vspace{-0.3cm}
\caption{{\small Comparison of predicted peculiar velocities along the line of sight for four selected galaxies.
The dashed line represents the IRAS model, the solid 
line shows our redetermination of the Tonry model, and the dotted 
line the TBAD00 model itself.}}
\label{fig:f5}
\end{figure}
\section{Application of the Velocity Models to the Cepheid Galaxy Sample}

In this section we apply the velocity models of \S 5 to the 
Cepheid galaxy sample discussed in \S 4. Our application is
limited to the twenty-seven galaxies listed
in Tables~\ref{tab:list} and~\ref{tab:d} that are not LG members.
As noted in \S 4, 26 of these 27 galaxies
have HST Cepheid data that have been
analyzed by the $H_0$KP team; the one galaxy for which ground-based
data are used, NGC 300, was analyzed by $H_0$KP team members
(Freedman \etal\ 1992)
and is thus expected to be on the same system.
The Cepheid sample used here thus constitutes
a uniform data set.

We now apply the \velmod\ method to the Cepheid data set, as we did in
\S 5 for the SBF data; here we determine $H_0.$ There is
now one key difference: the Cepheid galaxy distances $d$ are in Mpc,
and correspondingly, Eq.~(\ref{eq:pdr}) is rewritten \be P(\ln d|r) =
\frac{1}{\sqrt{2\pi}\Delta} \exp\left\{-\frac{\left[\ln(H_0
d/r)\right]^2} {2\Delta^2}\right\} \,.
\label{eq:pdr_ceph}
\ee
In the exercise of \S 5 a vector of free parameters,
$\bfp,$ on which the velocity model depended, was varied to
maximize likelihood. Now, $\bfp$ is held fixed at the values
determined in \S 5---i.e., the same velocity field is
used---and the only parameter that is varied to maximize likelihood
is $H_0.$ Aside from these differences, the \velmod\ procedure
applied to the Cepheid galaxies is identical to that applied
to the SBF data set.

\subsection{Results from the IRAS models}

To apply the IRAS model we must adopt a functional
form and value of $\sigma_v(r),$ the small-scale velocity
noise as a function of position. Following WS, we write
\be
\sigma_v(r) = \sigma_{v,0} + f_v \delta_g(r),
\label{eq:sigvr}
\ee 
where $\delta_g$ is the IRAS galaxy overdensity at position
$r$ along the line of sight. Thus, $\sigma_{v,0}$ is the
velocity noise in a mean-density environment, and $f_v$
represents the rate of increase of velocity dispersion
with density, an effect expected on theoretical grounds
and verified in N-body simulations (Kepner, Summers, \& Strauss 1997;
Strauss, Ostriker, \& Cen 1998). 
For the Cepheid sample we adopt $\sigma_{v,0}=135\,\kms$
and $f_v = 30\,\kms,$ similar to the values WSDK and WS found for
Tully-Fisher \velmod. As noted earlier, this value of $\sigma_v$
is considerably smaller than the $\sim 185\,\kms$ found for
the SBF sample, but the Cepheid galaxies are 
late-type spirals and are more likely to resemble
the Tully-Fisher galaxies in their dynamical properties.

There is one subset of galaxies within the Cepheid sample that are not
well-described by this model of $\sigma_v(r),$ namely, Virgo cluster
members. Virgo's relatively high ($\sim 650\,\kms$) velocity
dispersion cannot be matched by the linear increase with density at
the IRAS smoothing scale of $300\,\kms.$ To account for this we
``collapse'' Virgo, following WS and WSDK---i.e., we set the redshift
of each Virgo galaxy to its mean value of $cz_{{\rm LG,Virg}} =
1035\,\kms$ (Huchra 1985).  To account for uncertainty in this value,
we set the $\sigma_v=30\,\kms$ for each collapsed Virgo galaxy. The
Cepheid galaxies deemed likely Virgo members were NGC4536, NGC4321,
NGC4496A, NGC4535, and NGC4548. Their mean Cepheid distance is 14.7
Mpc. The distance of each one is within $\sim 1.5\sigma$ of this mean
value, each is  within 10\degs\ of the Virgo core at $\ell=283.8\degs,$
$b=74.5\degs,$ and each has a redshift within $700\,\kms$ of the Virgo
mean. Thus, these are all strong candidates for Virgo membership, and
collapsing them significantly improves the fit likelihood. However, as
we now show, whether or not we collapse Virgo makes little difference
to the derived value of $H_0.$

Table~\ref{tab:IRAS} presents the main results of applying the IRAS
velocity models to the Cepheid sample. Column 1 lists the value of
$\beta$  (the values of $\bfw_{{\rm LG}}$ used are always those
derived from the SBF fit at that value of $\beta,$ and
$\sigma_v(r)$ is always given as discussed in the previous
paragraph); Column 2 lists
the maximum likelihood value of $H_0$ and its $1\,\sigma$
uncertainty; columns 3 and 4 give the values
of the likelihood statistic ${\cal L}$ and a $\chi^2$
for the fit, which is discussed further below. Columns
5, 6, and 7 repeat the information of 3, 4, and 5 for
the case that Virgo is not collapsed, i.e., when the true
LG redshifts of the Virgo galaxies are used in the likelihood analysis. 

The most striking feature of Table~\ref{tab:IRAS} is the robustness
of $H_0$ with respect to changes in $\beta.$ Indeed, 
the results of the Table can be summarized
by the statement $H_0 = 85 \pm 2.5\,\kmsmpc$ at
65\% ($1\,\sigma$) confidence, irrespective
of the value of $\beta,$ provided it is in the range
$0.2$--$0.5$ that is allowed by the IRAS velocity
model applied to the SBF sample.\footnote{The application
of \velmod\ to Tully-Fisher samples by WSDK and WS
preferred higher values of $\beta,$ in the range
$0.4$--$0.6$ at 95\% confidence. We consider the SBF
result to be more reliable because of the greater precision
of SBF distances. In any case, we note that $\beta=0.4$ is
allowed by both the SBF and Tully-Fisher data, and thus
constitutes the preferred value at present.}
When Virgo is not collapsed, the likelihood statistics
and $\chi^2$ values indicate a much worse fit.
This occurs because the Virgo galaxy redshifts scatter
so widely about the mean cluster value, and our simple
model of velocity noise increase with density does
not account for this (possibly because the huge
central densities are smoothed out, and because
IRAS undercounts cluster cores). It is important to
note, however, that not collapsing Virgo affects the
derived value of $H_0$ very little, increasing it
by an amount about equal to the $1\,\sigma$ error estimate.
Thus our treatment of Virgo is not crucial to the
main conclusion of this paper.
\begin{deluxetable}{cccccccc}
\tablecolumns{8}
\tablewidth{0pc}
\tablecaption{Solutions for IRAS Velocity Fields\label{tab:IRAS}}
\tablehead{\colhead{} & \multicolumn{3}{c}{Virgo Collapsed} &\colhead{} &
  \multicolumn{3}{c}{No Virgo Collapse}  \\
 \cline{2-4} \cline{6-8} \\
\colhead{} & \colhead{$H_0$} & \colhead{} &  \colhead{} & \colhead{}
 & \colhead{$H_0$}  & \colhead{} &  \colhead{} \\
\colhead{$\beta$} & \colhead{($\kmsmpc)$} & \colhead{${\cal L}$} &  \colhead{$\chi^2$} & \colhead{}
 & \colhead{($\kmsmpc)$}  & \colhead{${\cal L}$} &  \colhead{$\chi^2$}}
\startdata 
$0.2$  & $84.9 \pm 1.6$ & $34.7$ & $44.7$ & & $85.6 \pm 2.4$ & $68.0$ & $79.9$ \\
$0.3$  & $84.6 \pm 1.9$ & $31.6$ & $38.8$ & & $86.4 \pm 2.5$ & $62.1$ & $72.8$ \\ 
$0.4$  & $85.3 \pm 2.2$ & $35.5$ & $39.4$ & & $87.5 \pm 2.5$ & $62.9$ & $74.8$ \\ 
$0.5$  & $86.2 \pm 2.8$ & $38.9$ & $39.3$ & & $88.6 \pm 2.8$ & $62.6$ & $77.6$ \\ 
\enddata
\end{deluxetable}

We discuss below the calculation of the
confidence intervals on $H_0.$ First, we describe the calculation
of the $\chi^2$ statistics listed in Table~\ref{tab:IRAS}.

\subsubsection{$\chi^2$ statistic for the velocity fits}

The likelihood statistic ${\cal L}$ 
does not by itself
provide a measure of goodness of fit. WSDK and WS pointed out
that it was not possible to construct a rigorous $\chi^2$ statistic
for their Tully-Fisher \velmod\ fits because the distance and
velocity errors were determined as part of likelihood maximization.
This is not the case here, however, because (i) we have reliable 
distance errors for the Cepheid galaxies, and (ii) we have fixed
the Cepheid velocity noise a priori (Eq.~\ref{eq:sigvr}). Thus
it makes sense to define and calculate a $\chi^2$
statistic to test the goodness of fit of our velocity models
to the Cepheid galaxy data.

We first define the expected distance in $\kms,$ or
{\em Hubble flow distance,} given
the redshift and the velocity model,
\be
E(r|cz) = \frac{\zinfint r^3 n(r) P(cz|r)\,dr}
{\zinfint r^2 n(r) P(cz|r)\,dr} \,,
\label{eq:ergcz}
\ee
along with the corresponding mean square distance
\be
E(r^2|cz) = \frac{\zinfint r^3 n(r) P(cz|r)\,dr}
{\zinfint r^2 n(r) P(cz|r)\,dr} \,
\label{eq:er2gcz}
\ee
and distance error,
\be
\delta r = \sqrt{E(r^2|cz) - \left[E(r|cz)\right]^2} \,.
\label{eq:Delw}
\ee
The Cepheid distance in $\kms$ is $H_0 d,$ where $d$ is the
Cepheid distance in Mpc and $H_0$ is the best-fit Hubble constant
for the model in question, and the corresponding error is
$H_0 d \Delta$ (see \S 5.1). The $\chi^2$ statistic measures
the difference between $H_0 d$ and $E(r|cz)$ in units
of the overall error:
\be
\chi^2 = \sum_{i=1}^{27} \frac{\left(E(r|cz) - H_0 d\right)^2}{(\delta r)^2 +
(H_0 d \Delta)^2 } \,.
\label{eq:defchi2}
\ee
It is important to note that the Hubble flow contribution to
the error, $\delta r,$ is not determined solely by the
velocity noise $\sigma_v.$ Rather, it is the integrated
effect of $\sigma_v$ along the line of sight, taking into
account the shape of the large-scale velocity field $u(r).$
In particular, in regions where $u'(r)<0$ the effect of
velocity noise is enhanced, creating comparatively large
$\delta r$ (see Appendix A of WS for further discussion).

The fourth and seventh columns of Table~\ref{tab:IRAS} list the $\chi^2$ values
for the Virgo-collapsed and uncollapsed fits. There are 26 degrees
of freedom for the fit---27 galaxies minus one free parameter---so
that the expected $\chi^2$ value is 26, with rms dispersion 
$\sqrt{52}=7.2.$ The Virgo-collapsed fits with $\beta\ge 0.3$ thus
have $\chi^2$ values within $2\,\sigma$ of their expected values.
These fits are therefore statistically acceptable, albeit with
a larger $\chi^2$ than desirable. 

Two points are worth
bearing in mind with regard to this statement, however. First,
the relatively high $\chi^2$ value is largely due to the influence
of one galaxy, NGC 1326A, which deviates from the fit
by $\sim 3\,\sigma.$ When this object is excluded the computed
$\chi^2$ is well within $1\,\sigma$ of the expectation.\footnote{Moreover,
NGC1326A has very little effect on the value of $H_0;$
for $\beta=0.4$ we find $H_0=84.2\,\kmsmpc$ when NGC1326A is
excluded.}
Second, the absolute value of $\chi^2$ is strongly
dependent on our adopted $\sigma_v(r).$ We chose $\sigma_v(r)$
similar to, though somewhat larger than, the value found
by WS for a Tully-Fisher sample. If we take $\sigma_v(r) 
= 185\,\kms,$ the value indicated by the SBF fit (\S 5.3),
we obtain $\chi^2=29.9$ for the $\beta=0.4$ fit,
fully compatible with expected value of 
$26 \pm 7.2.$ (The best-fit Hubble constant rises to $86.8\,\kmsmpc$
for this choice of $\sigma_v(r),$ a change of less than $1\,\sigma.$) 
We believe that the smaller $\sigma_v(r)$ is appropriate for
the late-type Cepheids, but this example shows that it
is difficult to assign unambiguous significance to our
$\chi^2$ statistic. However, the statement that 
{\em the SBF-constrained IRAS models
provide satisfactory fits to the 
Cepheid sample for\/} $\beta = 0.3$--$0.5$ is
a reasonable one in view of the above discussion. 
\begin{figure}[!ht]
\begin{center}
\includegraphics[scale=0.40]{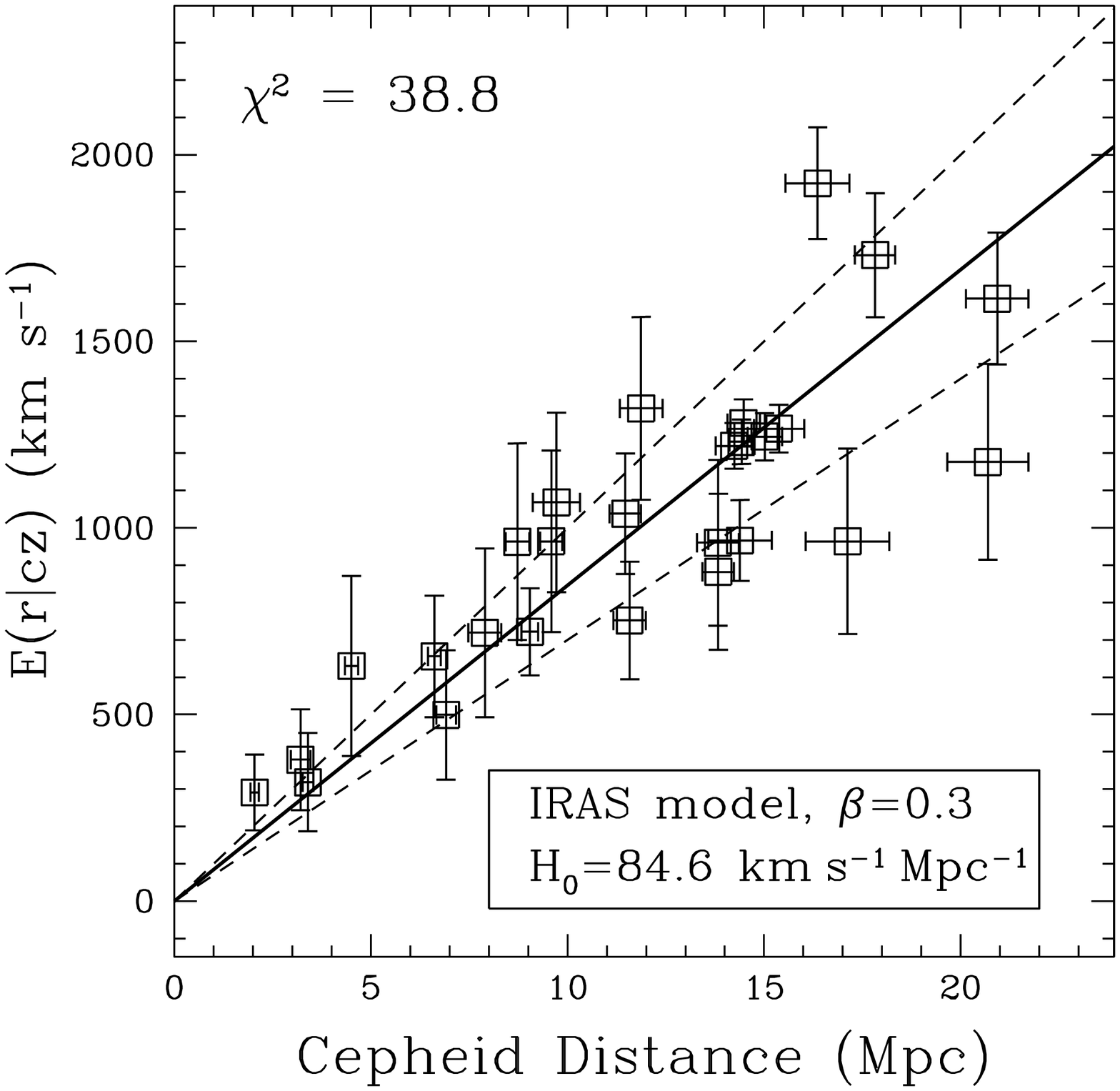}\hfill~
\includegraphics[scale=0.40]{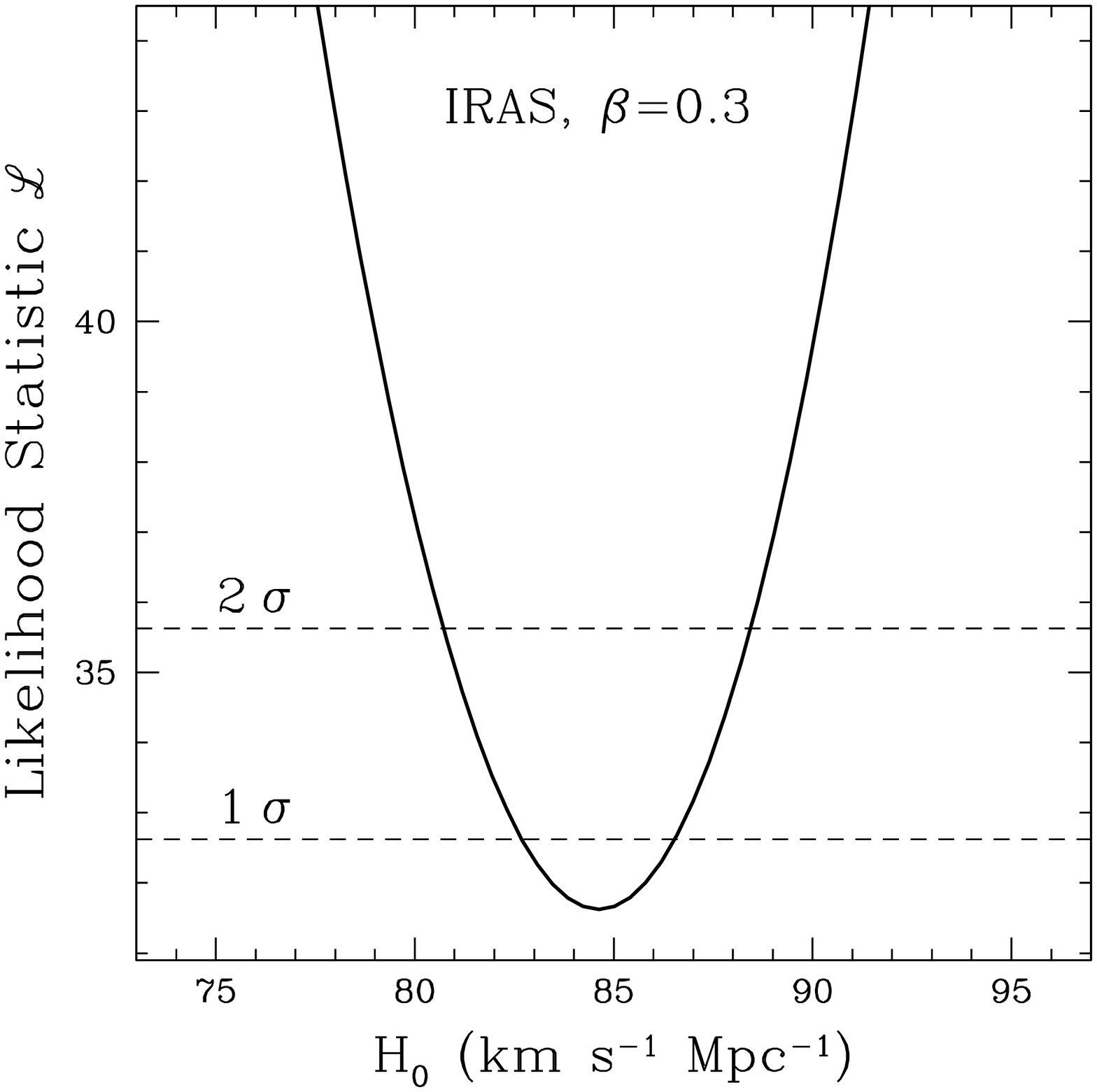}
\end{center}
\vspace{-0.3cm}
\caption{{\small Hubble diagram for 27 Cepheid galaxies.
The velocity model used is the IRAS $\beta=0.3$ linear
reconstruction; the expected distances in velocity
units, $E(r|cz),$ are plotted versus the Cepheid
distances in Mpc. The heavy solid line shows the best
fitting Hubble constant, $H_0=84.6\,\kmsmpc.$ The
flatter dashed line is $H_0=70\,\kmsmpc,$ and
the steeper dashed line is $H_0=100\,\kmsmpc.$}}
\label{fig:f6a}
\end{figure}

\begin{figure}[!ht]
\begin{center}
\includegraphics[scale=0.40]{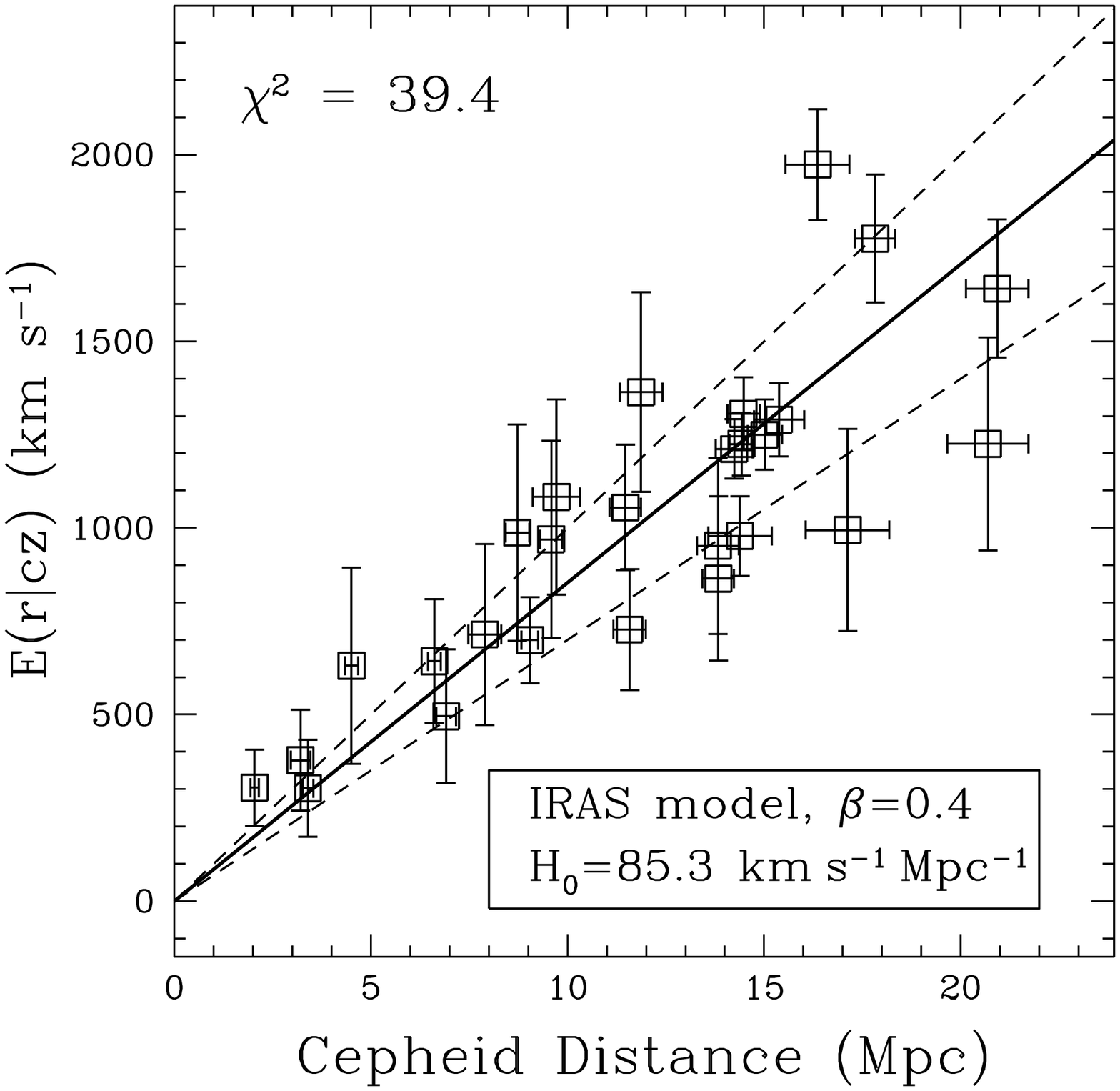}\hfill~
\includegraphics[scale=0.40]{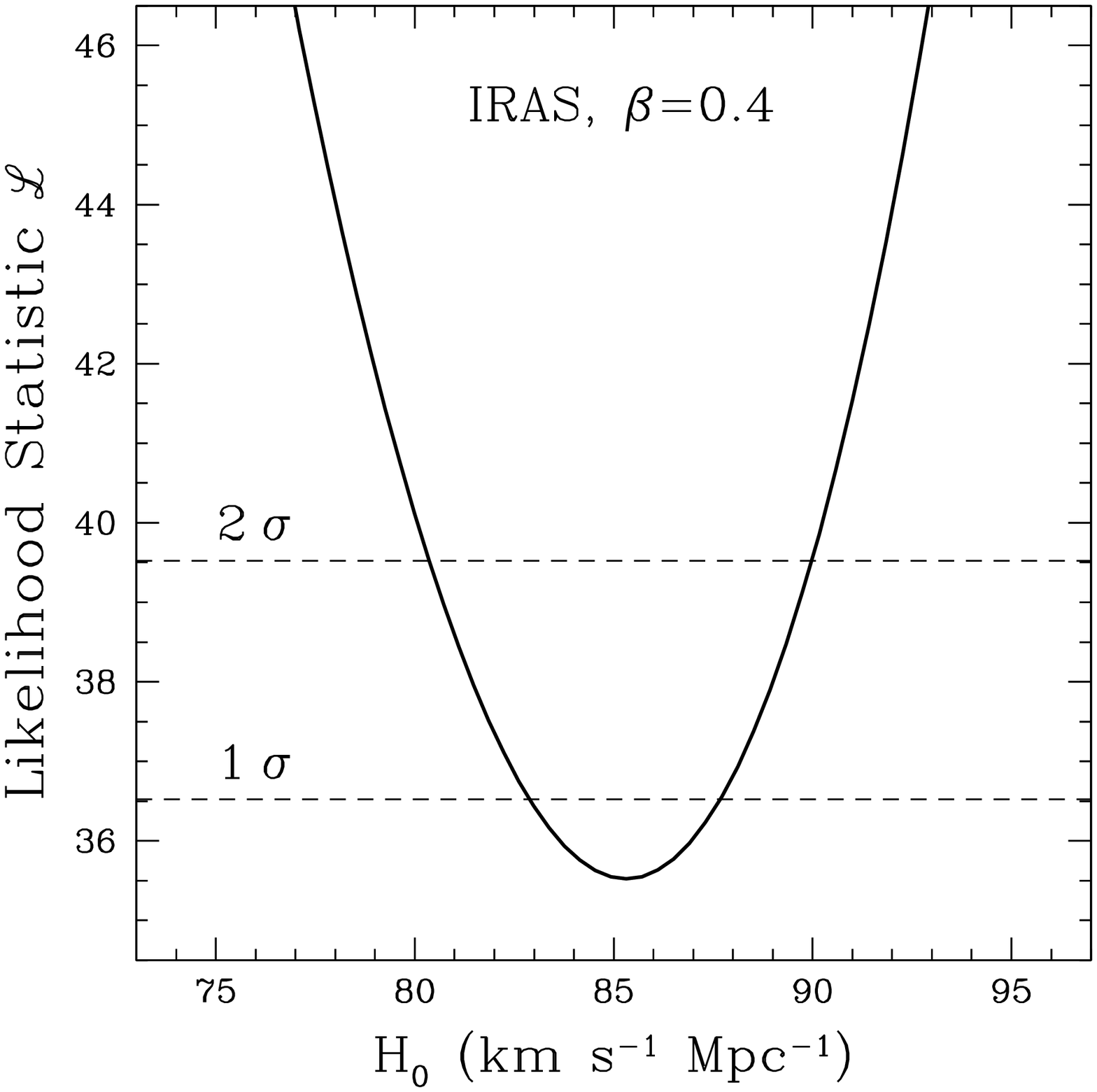}
\end{center}
\vspace{-0.3cm}
\caption{{\small Same as the previous figure, but for an IRAS
model with $\beta=0.4.$}}
\label{fig:f7a}
\end{figure}
 
\subsubsection{Hubble diagrams and confidence intervals}

To assess the quality of the fits and the derived values
of $H_0,$ we plot Hubble diagrams in the left hand
panels of Figures~\ref{fig:f6a} and~\ref{fig:f7a}. 
The abcissa is the quantity $E(r|cz),$ the expected value of
the Hubble flow distance given the LG frame redshift and the
IRAS velocity model, as defined by Eq.~(\ref{eq:ergcz}).
The vertical error bars are the quantity $\delta r$ defined
by Eq.~(\ref{eq:Delw}), and the horizontal error bars are
the Cepheid distance uncertainty $d\Delta.$ (We have approximated
all errors as being symmetric.) In each Hubble diagram we
plot the best-fit value of $H_0$ as a solid line through the
points, with dashed lines representing the values
$H_0=70\,\kmsmpc$ and $H_0=100\,\kmsmpc$ for comparison.

The diagrams validate the values of $H_0$ given in Table~\ref{tab:IRAS}. 
The solid lines represent a far better fit to the
majority of the data points than do the dashed lines
drawn for comparison. The cluster of points very near
the best fit Hubble line at a distance of 15 Mpc are
the five collapsed Virgo galaxies. Their error bars are
much smaller because we assume a velocity uncertainty of
only $30\,\kms$ for these objects, associated with
the uncertainty in the Virgo redshift. (The error bars are,
however, larger than $30\,\kms$ because of the interaction
between $\sigma_v$ and $u(r)$ mentioned above.) 
It is evident from the diagrams that the fits are not perfect,
with larger deviations being seen at the distance of
Virgo. This may represent significant departures from
the IRAS model in dense regions, a topic we further
discuss in \S 7.1. As noted above, however, our uncertainty
about the precise value of $\sigma_v(r)$ means we cannot
state with assurance that these points do not fit the model.

The Hubble constant errors we have quoted are calculated by
calculating the likelihood statistic ${\cal L}$ for a range
of values of $H_0$ near the best fit value. (We hold all
velocity field parameters, as well as $\sigma_v(r),$ constant
as we vary $H_0.$) As discussed in detail by WSDK, ${\cal L}$
has the property that, when one fit parameter is varied,
increases of $\Delta {\cal L} = \pm 1$ with respect to the
minimum yield the $1\,\sigma$ errors on the parameter,
while increases of $\Delta {\cal L} = \pm 4$ yield the
$2\,\sigma$ errors. This is strictly true if the likelihood
is Gaussian, or, equivalently, if the variation of
${\cal L}$ is parabolic near its minimum. This is
a good approximation in the present case. The right hand
panels of Figures~\ref{fig:f6a} and~\ref{fig:f7a}
plot ${\cal L}$ versus $H_0$ for the $\beta=0.3$ and $\beta=0.4$
models respectively. The horizontal dashed lines show
the $1$ and $2\,\sigma$ confidence intervals on $H_0$ as determined
by the procedure described above. These curves are the origin of
the 95\% confidence interval on $H_0$ of $5\,\kmsmpc$ quoted
in the Abstract.

\subsection{Results from the Tonry Model}

We apply the phenomenological
Tonry model to the Cepheid galaxies using the best-fit parameters arising
from the SBF velocity fit. We again collapse Virgo in the Cepheid fit.
We now fix the baseline velocity noise to $\sigma_v(r)=150\,\kms,$
allowing it to rise to the SBF values in the cores of the model's attractors.
The baseline noise is a bit higher than was the case for the IRAS
fit (Eq.~\ref{eq:sigvr}). 
However, the density dependence of $\sigma(r)$ for IRAS
was such that non-cluster galaxies have very nearly the same velocity
noise for the IRAS and Tonry model fits. The value of $H_0$ exhibits
slight sensitivity to the value of $\sigma_v,$ but the variation
is small relative to the statistical errors.

Figure~(\ref{fig:f8a}) shows the Hubble diagram resulting from
fitting the Tonry model to the Cepheid data (left panel), along
the likelihood versus $H_0$ curve (right panel). 
The maximum likelihood result find $H_0 = 91.8 \pm
1\,\kmsmpc$ at 65\% ($1\,\sigma$) confidence. This is considerably
larger, relative to the errors, than the result from the
IRAS model. However, note that the $\chi^2$ for the
Tonry model fit, indicated on the Figure, is substantially
larger than the value obtained for the IRAS fit.\footnote{The 
minimum value of the likelihood statistic ${\cal L}$ is also
markedly larger than for the IRAS fit; however, the absolute likelihood
statistics are not necessarily comparable for the two models owing
to differences in implementation, such as the adoption of
a constant density for the Tonry model.}  Indeed, it deviates
by more than $3\,\sigma$ from the expected value of 26. 
We conclude that the Tonry model is not an acceptable fit
to the Cepheid distances; a corollary is that the discrepancy
between the derived values of $H_0$ is not meaningful.

Note that the collapsed Virgo galaxies in the left panel of
Figure~\ref{fig:f8a} have vertical error bars that are much
larger than those for the IRAS fit. This reflects the very
strong Virgo infall inherent in the best-fit Tonry model---a
triple-valued zone is produced near Virgo, which causes a small
$\sigma_v$ to translate in to a large $\delta r.$ The large $\chi^2$
for the fit probably indicates that the Tonry model attributes
too much strength to the Virgo attractor as it attempts to
compensate for the missing mass concentrations and voids that the IRAS   
model contains.

Table~\ref{tab:plot} lists the data plotted in Figures~\ref{fig:f6a}, ~\ref{fig:f7a} and~\ref{fig:f8a}.

\begin{figure}[!ht]
\begin{center}
\includegraphics[scale=0.40]{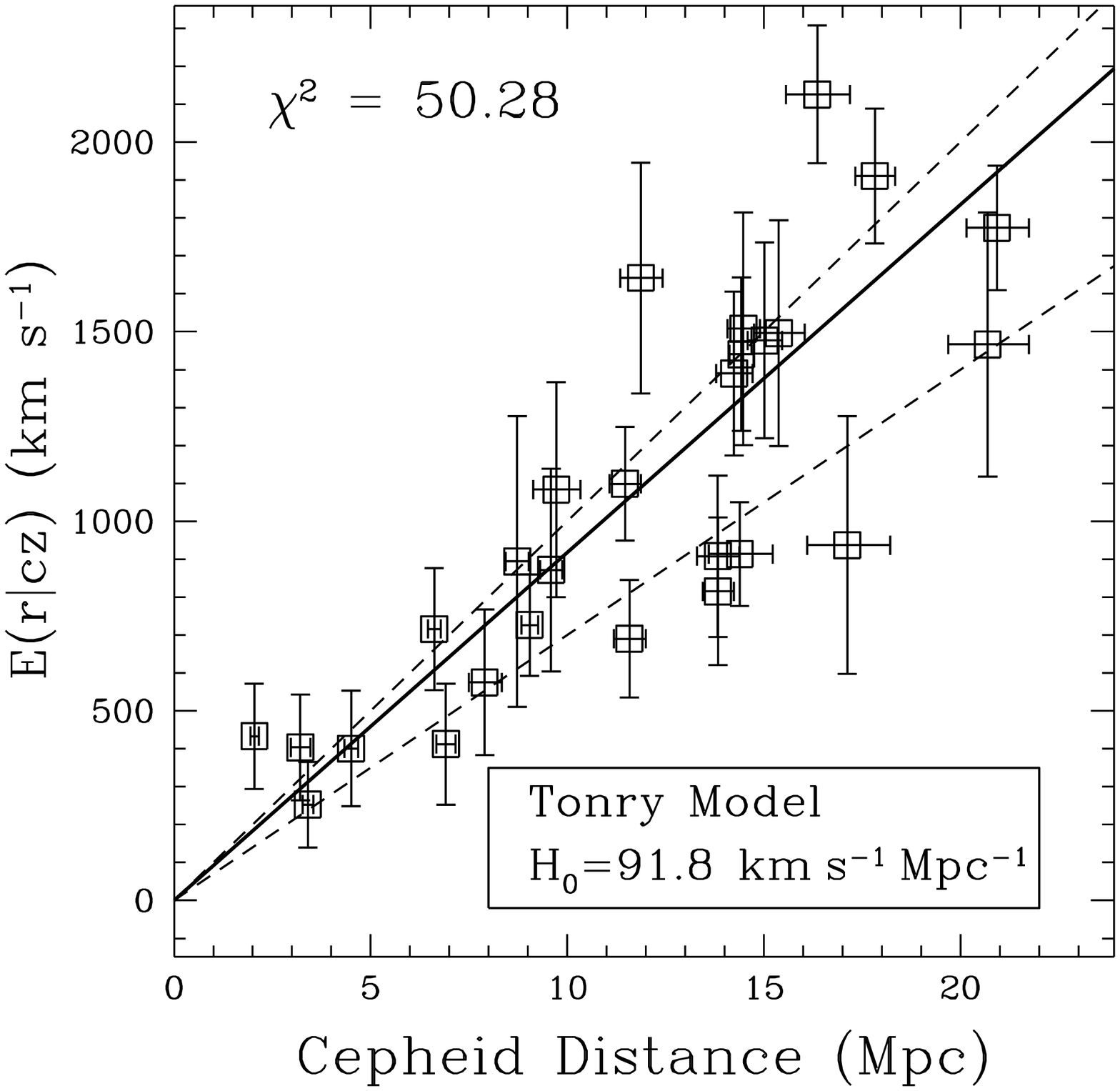}\hfill~
\includegraphics[scale=0.40]{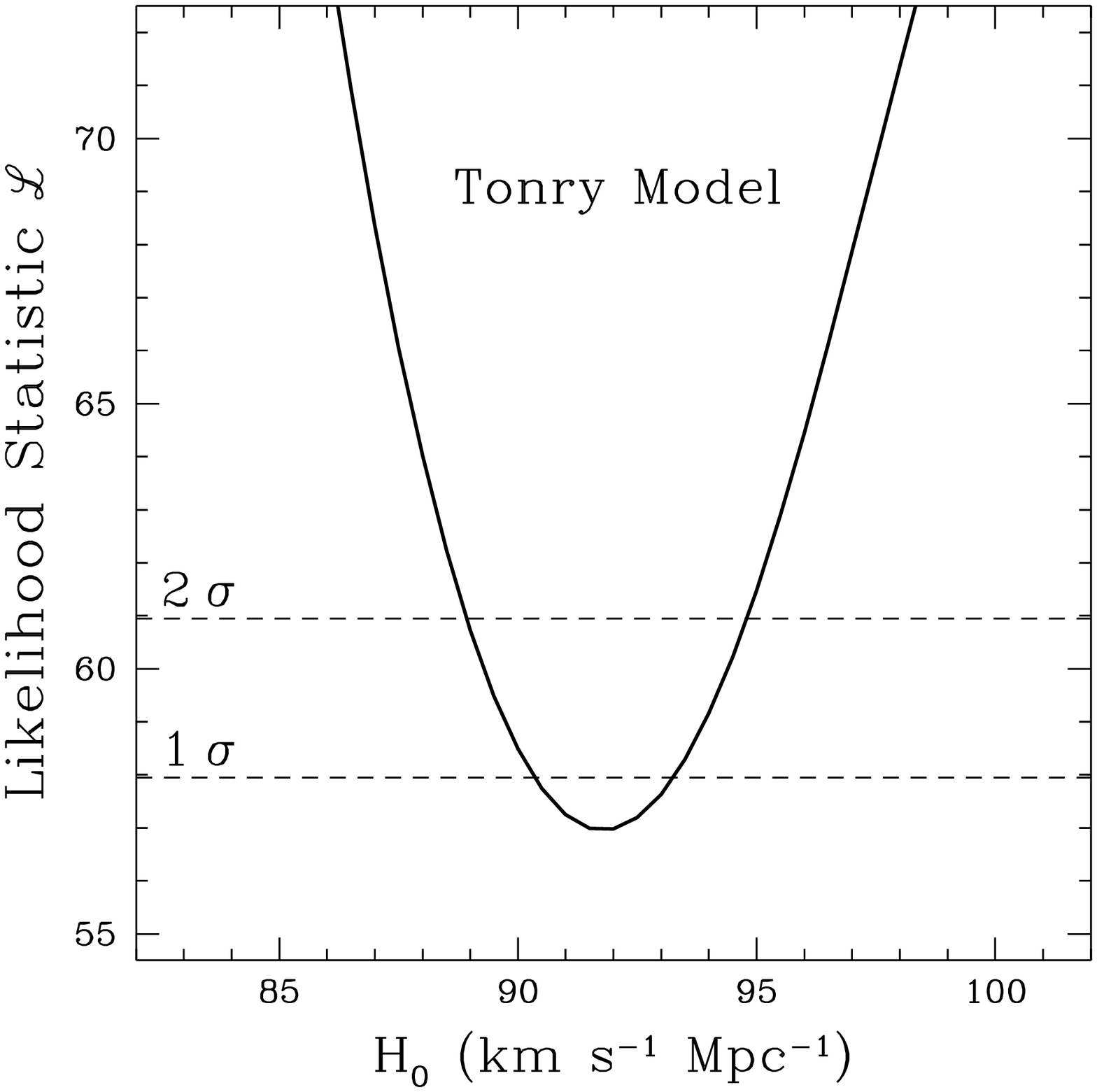}
\end{center}
\vspace{-0.3cm}
\caption{{\small Same as the previous figures, but for the phenomenological
Tonry model with $\sigma_v=135\,\kms$.}}
\label{fig:f8a}
\end{figure}

\begin{deluxetable}{lrrrrr}
\tablewidth{0pc}
\tablecolumns{6}
\tablecaption{Plotted Data\label{tab:plot}}
\tabletypesize{\footnotesize}
\tablehead{
\colhead{}&\colhead{} & \colhead{} &\colhead{$E(r|cz)$}&\colhead{$E(r|cz)$}&\colhead{$E(r|cz)$} \\
\colhead{}&\colhead{$d$}&\colhead{$cz_{{\rm CMB}}$}&\colhead{IRAS $\beta=0.3$}&\colhead{IRAS $\beta=0.4$}&\colhead{Tonry Model} \\
\colhead{Name} & \colhead{(Mpc)} & \colhead{(\kms)} & \colhead{$(\kms)$} & \colhead{$(\kms$)} &
\colhead{(\kms)}}
\startdata 
NGC 3031 & $ 3.40\pm 0.14$ & $ 46$ & $319
\pm 132$ & $325 \pm 122$ & $224 \pm 100 $ \\ NGC 300 & $ 2.04\pm 0.11$
& $ -89$ & $291 \pm 101$ & $206 \pm 74$ & $411 \pm 127 $ \\ NGC 5457 &
$ 6.91\pm 0.25$ & $ 360$ & $499 \pm 174$ & $549 \pm 164$ & $378 \pm
143 $ \\ NGC 5253 & $ 3.21\pm 0.25$ & $ 678$ & $379 \pm 135$ & $299
\pm 109$ & $369 \pm 133$ \\ NGC 4258 & $ 7.90\pm 0.42$ & $ 651$ & $719
\pm 226$ & $771 \pm 223$ & $378 \pm 126$ \\ NGC 4548 & $15.38\pm 0.64$
& $ 805$ & $1270 \pm 60$ & $1460\pm 60 $ & $541 \pm 169$ \\ NGC 925 &
$ 9.04\pm 0.21$ & $ 326$ & $722 \pm 116$ & $733 \pm 95 $ & $1500 \pm
300$ \\ NGC 2541 & $11.58\pm 0.41$ & $ 694$ & $752 \pm 157$ & $831 \pm
134$ & $675 \pm 142$ \\ NGC 3198 & $13.84\pm 0.39$ & $ 879$ & $882 \pm
209$ & $953 \pm 197$ & $793 \pm 178$ \\ NGC 4414 & $17.13\pm 1.06$ & $
988$ & $964 \pm 248$ & $1040 \pm 250$ & $866 \pm 312$ \\ NGC 3621 & $
6.62\pm 0.16$ & $1059$ & $656 \pm 163$ & $568 \pm 141 $ & $700 \pm
147$ \\ NGC 3627 & $ 8.73\pm 0.30$ & $1072$ & $963 \pm 263$ & $996 \pm
274 $ & $815 \pm 351$ \\ NGC 3319 & $13.83\pm 0.53$ & $ 978$ & $960
\pm 222$ & $1040 \pm 210$ & $884 \pm 195$ \\ NGC 3351 & $ 9.59\pm
0.28$ & $1123$ & $964 \pm 243$ & $988 \pm 244 $ & $821 \pm 235$ \\ NGC
7331 & $14.39\pm 0.82$ & $ 492$ & $966 \pm 109$ & $1010 \pm 80 $ &
$905 \pm 124$ \\ NGC 3368 & $ 9.72\pm 0.60$ & $1242$ & $1070 \pm 240$
& $1110 \pm 240$ & $1050 \pm 270$ \\ NGC 2090 & $11.47\pm 0.40$ &
$1002$ & $1040 \pm 160$ & $957 \pm 140 $ & $1090 \pm 135.3$ \\ NGC
4639 & $20.69\pm 1.03$ & $1328$ & $1180 \pm 262$ & $1250 \pm 270$ &
$1460 \pm 340$ \\ NGC 4725 & $11.87\pm 0.54$ & $1486$ & $1320 \pm 240$
& $1450 \pm 253$ & $1640 \pm 290$ \\ NGC 1425 & $20.93\pm 0.79$ &
$1413$ & $1610 \pm 180$ & $1560 \pm 170$ & $1770 \pm 150$ \\ NGC 4321
& $14.48\pm 0.41$ & $1893$ & $1280 \pm 60$ & $1480 \pm 60 $ & $1510
\pm 310$ \\ NGC 1365 & $17.83\pm 0.51$ & $1539$ & $1730 \pm 170$ &
$1690 \pm 150$ & $1900 \pm 170$ \\ NGC 4496A & $14.43\pm 0.30$ &
$2070$ & $1230 \pm 60$ & $1340 \pm 80 $ & $1440 \pm 200$ \\ NGC 4536 &
$14.24\pm 0.47$ & $2148$ & $1220 \pm 60$ & $1310 \pm 70 $ & $1390 \pm
220$ \\ NGC 1326A & $16.36\pm 0.81$ & $1730$ & $1920 \pm 150$ & $1900
\pm 130$ & $2120 \pm 170$ \\ NGC 4535 & $15.02\pm 0.43$ & $2293$ &
$1240 \pm 60$ & $1400 \pm 70 $ & $1470 \pm 260$ \\
\enddata
\end{deluxetable}

\section{Discussion and Summary}

We have argued in this paper that $H_0 = 85 \pm 5\,\kmsmpc$ at
95\% confidence, considering random error only. 
This result, if correct, leads to an expansion timescale $H_0^{-1} =
10.9$--$12.2$ Gyr, and thus an expansion age $t_0 = f(\omegam,\omegal)H_0^{-1}$
that is shorter still (see the discussion in \S 1), unless
$\omegam = 1-\omegal \simlt 0.25.$ For example, for 
$H_0=85\,\kmsmpc$ and an $\omegam=0.3,$ $\omegal=0.7$ cosmology,
$t_0 = 11.1$ Gyr. This expansion age may be compared with the
estimated age of the oldest globular clusters, $t_* = 12.8 \pm 1$ Gyr
($1\,\sigma$ uncertainty; Krauss 1999). 
At first blush, then, our estimated Hubble constant leads to a universe
younger than its oldest stars. Given this logical contradiction, our
result obviously requires further scrutiny. We discuss 
a number of salient issues in this final section.

\subsection{Why do we disagree with the $H_0$KP?}

The $H_0$KP team reported $H_0=71 \pm 6 \, \kmsmpc$ (Mould \etal\ 2000).
However, of their reported 9\% ($1\,\sigma$) error, approximately
6.5\% is systematic error due mainly to uncertainty in the distance
to the LMC. This systematic error
affects our value in precisely the same way as theirs,
and thus should not be considered in comparing our $H_0$ estimates.
The $1\,\sigma$ {\em random\/} error in the $H_0$KP Hubble constant
is $\sim 4.4\,\kmsmpc,$ corresponding to a $2\,\sigma$ error
of $\sim 9\,\kmsmpc.$ Thus, the $H_0$KP estimate overlaps with
ours only at the very edges of our respective $2\,\sigma$ error bars
(i.e., at $80\,\kmsmpc$). Since we have used the same Cepheid
data set to arrive at our estimates, and therefore share much of
their random error, this constitutes a significant
disagreement. 

There are two principal causes of this disagreement.  The first is the
difference in Cepheid calibration as discussed in \S 3. Our OGLE-based
PL calibration produces distances that are smaller by $\sim 5\%$ on
average, primarily because the OGLE calibration yields larger
reddening and thus larger extinction estimates. When applied to the
$H_0$KP distances and their procedure for estimating the Hubble constant,
the OGLE calibration should bring their value up to $\sim
75\,\kmsmpc,$ closer to the value we derive, though still
inconsistent, given that we use the same data.

The second source of disagreement is more fundamental:
the different strategies we have adopted for determining $H_0.$
The $H_0$KP approach (referred to as ``Method I'' in \S 2) 
has been to use the Cepheid
galaxies as calibrators for secondary distance indicators (DIs),
especially Type Ia Supernovae (SN Ia), the Tully-Fisher and
Fundamental Plane relations, and the SBF relation. The
secondary DIs are then applied to galaxies at much
larger distances than the Cepheid galaxies themselves,
typically in the $3000$--$10,000\,\kms$ range,
where peculiar velocities can be largely neglected.
In contrast, we have derived $H_0$ from the Cepheid galaxies
themselves, correcting the effects of non-Hubble motions
with velocity field models (``Method II'' of \S 2).

The two strategies are subject to different pitfalls. Method I can go
awry with the propagation of Cepheid errors into the calibration of
the secondary DIs.  Such errors might occur for a variety of reasons,
the foremost being the difficult nature of the measurements. SN Ia are
often historical, i.e., occurred many years or decades ago, and the
data on their brightnesses may not be consistent with modern
methods. And yet, such historical SN Ia must be used in the
calibration procedure, SN Ia being rare events and Cepheid galaxies
being few in number. Four out of six SN Ia calibrated by Gibson \etal\
(2000) occurred prior to 1990, and two of these occurred prior to
1975. This small sample of calibrators introduces the biggest
uncertainty in the SN-based $H_0$ (Suntzeff \etal\ 1999). The
calibration of the SBF and Fundamental Plane (FP) methods suffer from
another problem: Cepheids are found in late-type spiral galaxies,
whereas SBF applies best, and FP applies only, to early type galaxies.
Consequently, the absolute calibration of the FP relation using
Cepheid distances (Kelson \etal\ 2000) must be obtained indirectly, by
assuming that the Cepheid calibrators and the FP ellipticals are
members of a group lying at a common distance.  The SBF relation has
been calibrated using a small number of spirals with prominent bulges
(Ferrarese \etal\ 2000a), but possible stellar population differences
between spiral bulges and ellipticals make the validity of this
calibration uncertain (Tonry \etal\ 1997; TBAD00).

Neither of the above problems applies to the $H_0$KP calibration
of the Tully-Fisher relation by Cepheid galaxies (Sakai \etal\ 2000). 
However, Sakai \etal\ obtained their value of $H_0$ 
from a single $I$ band Tully-Fisher data set, that of Giovanelli and
collaborators (e.g., Giovanelli \etal\ 1997). Although this data set
is of high quality, there are a number of other large Tully-Fisher
data sets of recent vintage that were not considered by Sakai \etal, such
as those collected in the the Mark III Catalog (Willick \etal\ 1997a).
Tully-Fisher measurements are prone to systematic differences in
velocity width and photometric measurement conventions, and application
of the Cepheid-calibrated Tully-Fisher relation to a wider
range of Tully-Fisher data sets is needed; an important first step in this
direction has been taken by Tully \& Pierce (1999).

A coincidence of multiple secondary DI miscalibrations is very unlikely,
at best. Nevertheless, our Method II analysis does not involve
secondary DI's and cannot suffer from the problem of propagated
Cepheid calibration errors.\footnote{It does, of course, potentially
suffer from calibration error in the Cepheid PL relation, but this is
essentially the problem of the LMC distance.} However, our approach is
vulnerable to inaccurately modeled peculiar velocities because we
measure $H_0$ locally (see the discussion in \S 2).  Our peculiar
velocity models are ``state-of-the-art,'' especially the IRAS models,
and we have optimized them with respect to the SBF data set, which is
the best current sample for constraining the local velocity field.  It
is evident from our Hubble diagrams, however, that our velocity models
are not perfect. A clear indication of this is the ridge of $\sim 5$
galaxies that lie well below our best-fit Hubble line in
Figures~\ref{fig:f6a} and~\ref{fig:f7a}. These are objects that lie
within, and in the background of, the Virgo-Ursa Major region, and
that are falling in toward Virgo or Ursa Major at higher velocity than
predicted by the model.  None of these objects deviates from our model
at more than the $2\,\sigma$ level---indeed, the only $3\,\sigma$
deviant point is NGC 1326A, which is above the $H_0=100\,\kmsmpc$ line
at a distance of 16.4 Mpc---but it is still disquieting to see the
large scatter at distances near and beyond Virgo. This scatter does
not invalidate our approach, but reminds us that peculiar velocities
are not fully accounted for in our model, and that some caution is
needed in interpretation. 

\subsection{Considerations for future work}

The discussion above shows that the debate on the Hubble
constant will continue.  Its ultimate resolution will
require that several key needs are met:
\begin{enumerate}
\item {\it More nearby galaxies with accurate Cepheid
distances}.--- Our local measurement of $H_0$ could be greatly improved
with a larger, and more uniformly distributed, sample of Cepheid
galaxies. It is to be hoped that further observations by the HST, and
later by NGST, will yield such samples. Indeed, such measurements
would enable a better grasp of the systematic error in Method II
values of $H_0$, along the lines of an angular variance analysis
suggested by Turner, Cen \& Ostriker (1992).

\item {\it Improved models for the local velocity
field}.---  The IRAS model presented here is a reasonable but imperfect
fit to both the SBF and the Cepheid distances. Work is currently under
way by WNSB to test enhancements of the model. In particular, we
intend to further investigate the effects of nonlinear dynamics and
nonlinear bias.  Another dynamical variable that needs to be better
understood is the small-scale velocity noise and its dependence on
galaxy density. If these efforts succeed, the uncertainty in $H_0$ due
to peculiar velocities will be reduced.

\item {\it A re-calibration of secondary DIs}.--- As noted in \S 2, both
the ``distant'' and ``local'' strategies for measuring $H_0$ should be
pursued, and eventually they should agree.  We have pointed to a few
issues where the $H_0$KP calibrations of secondary DIs could contain
subtle errors. For further investigation, one could use the Cepheid
distances to calibrate the Tully-Fisher relations for samples not
considered by Sakai \etal\ 2000, in particular, those tabulated in the
Mark III Catalog (Willick \etal\ 1997a), especially after possible
calibration errors in that and other catalogs have been corrected via
comparison with the uniform, all-sky Shellflow survey recently
presented by Courteau \etal\ (2000).
 
\item {\it Further exploration of the ``Hubble bubble''}.--- The approach
of this paper, and, to a lesser extent, that of the $H_0$KP, could
overestimate the Hubble constant if the local universe ($d \simlt
30\hmpc$) is expanding more rapidly than the global average, as has
been suggested by Zehavi \etal\ (1998) on the basis of SN Ia data
within $\sim 10,000\,\kms.$ Such a situation is certainly possible on
theoretical grounds; a fractional mass fluctuation $\delta_M$ within a
sphere of radius $R$ produces a deviation of the Hubble constant
within that sphere of $\delta H_0/H_0 \approx - \omegam^{0.6} \delta_M
/ 3$ (see Turner, Cen \& Ostriker 1992 and Tomita 2000 for more
detailed theoretical analyses). Typical mass fluctuations on scales $R
\simgt 10\hmpc$ are $\simlt 1$ in most cosmological scenarios, so that
one would expect that the local value of $H_0$ on a $\sim 10\,\hmpc$
scale to deviate by $\sim 10$--20\% from the global value if $\omegam
\approx 0.3,$ with smaller deviations on larger scales. However, for
our local value of $H_0$ to {\em exceed\/} the global value, it would
be necessary for our local neighborhood to be {\em underdense\/}
relative to the mean. And yet, the 1.2 Jy IRAS density field suggests
that the opposite is true---our local region within 20 Mpc is
overdense relative to the volume within 100 Mpc well that is
well-sampled by IRAS.  Moreover, recent Tully-Fisher data (Dale \&
Giovanelli 2000) do not support the claim of Zehavi \etal\ (1998) for
a local Hubble bubble.  It is prudent to view this issue as an open
one for now, and to continue to test the relationship between the
local and distant Hubble flow. Such tests do not require DIs that are
absolutely calibrated, and thus are independent of the question of
$H_0$ itself.

\item {\it More accurate absolute calibration of the PL relation}.--- As
discussed in \S 2, the largest systematic error in the analysis of
this paper is due to uncertainty in the zero point of the Cepheid PL
relation, which is itself due to uncertainty in the distance to the
LMC.  We have adopted the ``canonical'' value $\mu_{{\rm LMC}}=18.50$
in this paper, the same as that adopted by the $H_0$KP. Our $H_0$
estimate could be range between $78$ and $98\,\kmsmpc$ depending on
the distance to the LMC chosen.

\end{enumerate}

It is, finally, worth taking a moment to consider the
question, What if $H_0$ really is as large as, say,
$90\,\kmsmpc?$ Would that constitute a ``crisis for
Big Bang cosmology,'' a claim that has been heard in some quarters?
The answer, for the moment, is clearly ``No,'' for two reasons.
First, the globular clusters could still be as young
as $t_*=10$ Gyr. Such a young age is unlikely but possible
at the few percent level (Krauss 1999).
If this were the case, then $t_0 > t_*$ for $H_0=90\,\kmsmpc$
and $\omegam=0.3,$ $\omegal=0.7.$ One would then require only
that the globular clusters formed very shortly ($\simlt 10^8$ yr)
after the Big Bang, which is not impossible.
Second, even if $t_*=13$ Gyr is correct,
one can obtain $t_0 \ge t_*$ for $H_0=90\,\kmsmpc$
if $\omegam = 1-\omegal \le 0.13.$ A universe of
such low density has not been ruled out. In short, 
a Hubble constant $\approx 90\,\kmsmpc$ does not pose
an insurmountable problem for Big Bang cosmology
so long as the ages of the oldest stars and the values
of the parameters $\omegam$ and $\omegal$ remain poorly
determined.  

\subsection{Summary}

We have presented a new determination of the Hubble constant using
Cepheid PL data published by the $H_0$KP. Rather than use the nearby ($d
\simlt 20$ Mpc) Cepheid galaxies as calibrators for secondary distance
indicators, which are then applied to more distant ($d \simgt 50$ Mpc)
galaxies for which peculiar motions are fractionally small (the $H_0$KP
strategy), we use Cepheid galaxies directly to measure $H_0.$

We first redetermined the Cepheid galaxy distances using a calibration
of the PL relation derived from a large sample of LMC Cepheids
presented by the OGLE group. Our absolute PL calibration assumed
$\mu_{{\rm LMC}}=18.5.$ (We reemphasize that the $H_0$KP group will
shortly present their own revision of Cepheid distances in light of
the OGLE LMC Cepheid data [Madore \& Freedman 2000, in preparation;
Freedman \etal\ 2000, in preparation].)  We then presented two models
of the local peculiar velocity field.  The first was obtained from the
IRAS galaxy density field using the linear relation between
large-scale velocity and density fields and the assumption that IRAS
galaxies trace the mass density field up to a linear biasing factor
$b.$ The IRAS model applies in the Local Group reference frame.  The
second was the phenomenological model of TBAD00, which applies in the
CMB reference frame.  The Tonry model assumes the local velocity field
is dominated by infall to the Virgo and Great Attractors, along with a
dipole and quadrupole term. We optimized each model by fitting it,
using the maximum likelihood \velmod\ algorithm of WSDK and WS, to a
calibration-free 281-galaxy subset of the SBF sample of Tonry \etal\
(2001), currently the most accurate set of relative-distances for
galaxies in the nearby ($cz \simlt 3000\,\kms$) universe.  In the case
of the IRAS model, this optimization consisted mainly of determining
the value of $\beta=\omegam^{0.6}/b,$ which was found to be $0.38 \pm
0.06$ ($1\,\sigma$ error), with $0.2 \le \beta \le 0.5$ allowed at the
$3\,\sigma$ level.  For the Tonry model the optimization involved
constraining several parameters that determine the influence of the
Virgo and Great Attractors. The velocity model fits used only relative
distances for the SBF galaxies and thus in no way prejudiced our
subsequent determination of $H_0.$

We then applied the IRAS and Tonry velocity models to 27 Cepheid
galaxies, again using the \velmod\ algorithm, with the one remaining
free parameter now being the Hubble constant.  This yielded $H_0=85
\pm 5\,\kmsmpc,$ essentially independent of the value of $\beta,$ when
the IRAS velocity field was used. When the Tonry model was used we
obtained $H_0 = 92 \pm 5\,\kmsmpc.$ The quoted random errors are at
the 95\% confidence level. The IRAS model produced a better fit
likelihood than the Tonry model, and a Hubble diagram with markedly
less scatter. We thus favor the result from IRAS, and adopt the IRAS
value of $H_0$ for our final conclusion.  This value is significantly
larger than the $H_0$KP result, $H_0=71 \pm 6\,\kmsmpc$ (Mould \etal\
2000). 

Until Method I and Method II analyses give consistent results,
we find it untenable to state that $H_0$ is known to within $10
\%$. We discussed at length in \S 7.1 several possible reasons for the
difference, as well as (\S 7.2) a number of lines of further
investigation needed to clarify the issue. Should the larger value we
quote here turn out to be correct, it would be difficult to reconcile
the expansion age of the universe, $t_0 = 11.1 $ Gyr for
$\omegam=0.3,$ $\omegal=0.7,$ with the estimated ages of oldest
globular clusters, $12.8 \pm 1$ Gyr (Krauss 1999). However, the
remaining uncertainty in these stellar ages, as well as in the
cosmological density parameters, is sufficiently large that a Hubble
constant as large or even somewhat larger than what we have argued for
here does not yet pose a logical inconsistency for Big Bang cosmology.

\acknowledgements 

Sadly, Jeff died in a car accident only two days
after we (PB and JAW) first huddled over the APJ referee's report. I'd
like to think that some part of Jeff's wisdom shows in my own halting
revisions to this paper.  I would like to thank Stephane Courteau \&
Michael Strauss for helping with the completion of this work; I am
also grateful for Tod Lauer's referee's report.  I would have liked to
thank JAW for his time---so valuable to me now. Jeff's original
acknowledgements follow below.

This paper would not have been possible without the generous
assistance of a number of individuals who shared data and expertise.
We are particularly grateful to Wendy Freedman for timely input
regarding the evolving calibration of the PL relation and helpful
comments on an initial draft of this paper. $H_0$KP team members Laura
Ferrarese and Brad Gibson also provided useful input on calibration
issues. In addition, we acknowledge the entire $H_0$KP team for their
epochal achievement in putting together the HST Cepheid database.
Pierre Lanoix is thanked for his efforts in assembling and maintaining
the Extragalactic Cepheid Database.  Jeff Newman kindly provided his
Cepheid data for NGC 4258 in advance of publication. Special thanks
are due to Vijay Narayanan and Michael Strauss for producing
IRAS-predicted peculiar velocities and densities used in this paper,
and for helpful comments on the paper.  Finally, we owe a large debt
to John Tonry and his collaborators for their fine SBF data set, which
enabled us to constrain the local velocity field. We are particularly
grateful to John Blakeslee who shared the SBF data, and his wisdom on
how to use it, with us prior to its publication.
  
JAW was supported by a Cottrell Scholarship of Research Corporation
and a Terman Fellowship from Stanford University, and, during the
initial phase of this work, NSF grant AST96-17188.

\appendix

\section{Calculation of random distance errors}

Although the systematic zero point errors
in the Cepheid PL relation dominate the distance error
budget, the weights we assign to the galaxies in the Hubble
constant fit must be determined by random error only.
To calculate this error we must account for
how PL scatter couples with the reddening determination.
The result of this coupling is that the distance error
is markedly larger than one might naively expect.
 
\newcommand{\dmui}{\delta\mu_I}
\newcommand{\dmuv}{\delta\mu_V}
We denote the I and V band galaxy (random) distance modulus
errors $\dmui$ and $\dmuv.$ These errors
are due to PL scatter, and we assume them to be distributed
as Gaussian random variables with mean zero and rms dispersions
$\sigma_I/\sqrt{\nceph}$ and $\sigma_V/\sqrt{\nceph},$
where $\sigma_I$ and $\sigma_V$ are the I and V band
rms PL scatter. The modulus errors induce an error
$\delta (V-I) = \dmuv-\dmui$ in the mean $(V-I)$ color, which
leads in turn to an error in the assumed reddening given by
\be
\delta E(B-V) = \frac{\delta (V-I)}{R_V - R_I} \,,
\ee
where the $R_X\equiv A_X/E(B-V)$ are given in Table~\ref{tab:OGLE}. 
The corresponding error in the distance modulus
we assign to the galaxy is 
\be
\delta\mu^{red} = -\frac{1}{2}\,(R_V+R_I)\, \delta E(B-V) = -\frac{1}{2}\,\frac{R_V+R_I}{R_V-R_I}
\,\delta (V-I) \,,
\label{eq:dmured}
\ee
where we have assumed that the V and I band data are equally weighted
in the distance determination, as they are in
the present application. The negative sign in Eq.~(\ref{eq:dmured}) arises
because PL errors that make the galaxy appear redder result in its
being {\em overcorrected\/} for extinction, i.e., assigned too small a
distance modulus. 

Eq.~(\ref{eq:dmured}) represents only that part of the distance modulus error
due to reddening determination error. To it we must add the
direct error due to PL scatter, $\delta\mu^{direct}=(\dmuv+\dmui)/2.$
Adding the two, and writing the result in terms of the independent
variables $\dmuv$ and $\dmui,$ we obtain the overall galaxy distance
modulus error,
\be
\delta\mu=\frac{1}{2}\left[(1-\Delta_f)\,\dmuv+(1+\Delta_f)\,\dmui\right] \,,
\label{eq:dmu}
\ee
where we have defined
\be
\Delta_f \equiv \frac{R_V + R_I}{R_V - R_I} = 4.0625 \,.
\ee
The numerical value of $\Delta_f$ 
follows from the values of $R_V$ and $R_I$ in Table~\ref{tab:OGLE}.
Eq.~(\ref{eq:dmu}) tells us that $\delta\mu$ is normally distributed with mean
zero and rms dispersion
\be
\sigma_\mu = \sqrt{\left(\frac{1-\Delta_f}{2}\right)^2 \sigma_V^2/\nceph+
\left(\frac{1+\Delta_f}{2}\right)^2\sigma_I^2/\nceph} \,.
\ee
If we furthermore take $\sigma_V \approx \sigma_I =
\sigceph,$ the last equation reduces to
\be
\sigma_\mu = \frac{\sigceph}{2\sqrt{\nceph}}\,\sqrt{(1-\Delta_f)^2+
(1+\Delta_f)^2} = 2.96\,\frac{\sigceph}{\sqrt{\nceph}} \,.
\label{eq:sigmu}
\ee
The corresponding expression for the rms error in the
mean reddening for the galaxy is
\be
\sigma_{E(B-V)} = \frac{\sqrt{2}}{R_V - R_I}\frac{\sigceph}{\sqrt{\nceph}}
= 1.1\,\frac{\sigceph}{\sqrt{\nceph}} \,.
\label{eq:sigebv}
\ee
We use Eqs.~(\ref{eq:sigmu}) and~(\ref{eq:sigebv}), with $\sigceph=0.15$ mag, to
calculate the distance modulus and reddening errors
in columns 2 and 4 of Table~\ref{tab:d}. (The distance errors
in column 3 are computed directly from the
modulus errors in the standard fashion.)

\end{document}